\documentclass[12pt]{article}
\usepackage{graphicx}
\usepackage{xcolor}

\usepackage{scicite}

\usepackage{times}
\usepackage{amsmath}
\usepackage{mathtools}

\newenvironment{sciabstract}{%
\begin{quote} \bf}
{\end{quote}}

\newcounter{lastnote}

\newcommand{\ie}{\textit{i.e. }}
\newcommand{\eg}{\textit{e.g. }}

\title{Fitts' Law for speed-accuracy trade-off describes a diversity-enabled sweet spot in sensorimotor control}

\author
{Yorie Nakahira$^{1,\#}$, Quanying Liu$^{1,2\#}$, Terrence J. Sejnowski$^{3,4\ast}$, John C. Doyle$^{1\ast}$\\
\\
\normalsize{$^{1}$Division of Engineering and Applied Science, California Institute of Technology,}\\ \normalsize{Pasadena, CA 91125, USA}\\
\normalsize{$^{2}$Department of Biomedical Engineering, Southern University of Science and Technology,}\\ \normalsize{Shenzhen, 518055, China}\\
\normalsize{$^{3}$The Salk Institute for Biological Studies, La Jolla, CA, USA}\\
\normalsize{$^{4}$Division of Biological Sciences, University of California, San Diego, La Jolla, CA, USA}\\
\normalsize{$^{\#}$These authors contributed equally}\\
\normalsize{$^\ast$To whom correspondence should be addressed; E-mail: doyle@caltech.edu, terry@salk.edu.}
}

\date{}

\begin{document} 


\baselineskip 20pt


\maketitle


\begin{sciabstract}
Human sensorimotor control exhibits remarkable speed and accuracy, and the tradeoff between them is encapsulated in Fitts’ Law for reaching and pointing.  While Fitts related this to Shannon’s channel capacity theorem, despite widespread study of Fitts’ Law, a theory that connects implementation of sensorimotor control at the system and hardware level has not emerged. Here we describe a theory that connects hardware (neurons and muscles with inherent severe speed-accuracy tradeoffs) with system level control to explain Fitts’ Law for reaching and related laws. The results supporting the theory show that diversity between hardware components is exploited to achieve {\em both} fast and accurate control performance despite slow or inaccurate hardware. Such ``diversity-enabled sweet spots'' (DESSs) are ubiquitous in biology and technology, and explain why large heterogeneities exist in biological and technical components and how both engineers and natural selection routinely evolve fast and accurate systems using imperfect hardware.

\end{sciabstract}


\newpage

Human sensorimotor control is remarkably fast and accurate despite its implementation with slow or inaccurate components~\cite{todorov2002optimal,nagengast2011risk,franklin2011computational,lac1995learning,lisberger2010visual,sterling2015principles}. The trade-off between speed and accuracy is quantified in Fitts' Law for reaching (tasks such as finding a target with eye gaze, hand, or mouse): The time required for reaching a target of width $W$ at distance $D$ scales as $log_2(2D/W)$  \cite{fitts1964information,wiki}. The logarithmic relation between reaching time and target width allows faster speed with only a small decrement in accuracy. On the other hand, the speed-accuracy tradeoffs (SATs) of hardware implementing the control for reaching can be much more severe: Improved speed or accuracy in nerve signaling or muscle actuation requires profligate biological resources~\cite{sterling2015principles}, and as a consequence, very few nerve and muscle types are both fast and accurate~(Fig.~\ref{figure:hardware-SAT}). This apparent discrepancy between speed-accuracy tradeoffs in sensorimotor control vs.  neurophysiology raises the question: How does nature compensate for neurophysiological hardware constraints in sensorimotor control?

In this paper, we address this question with a networked control system model that relates SATs in sensorimotor control and neurophysiology. The model characterizes how hardware SATs in nerves and muscles impose fundamental limits in sensorimotor control, with Fitts' Law as a special case. The results show that appropriate speed-accuracy \textit{diversity} at the level of neurons and muscles (hardware) facilitates speed and accuracy in neurologic control performance despite the slow or inaccurate hardware, yielding a ``diversity-enabled sweet spot.''

\begin{figure}
\centering
\includegraphics[width=0.3\textwidth]{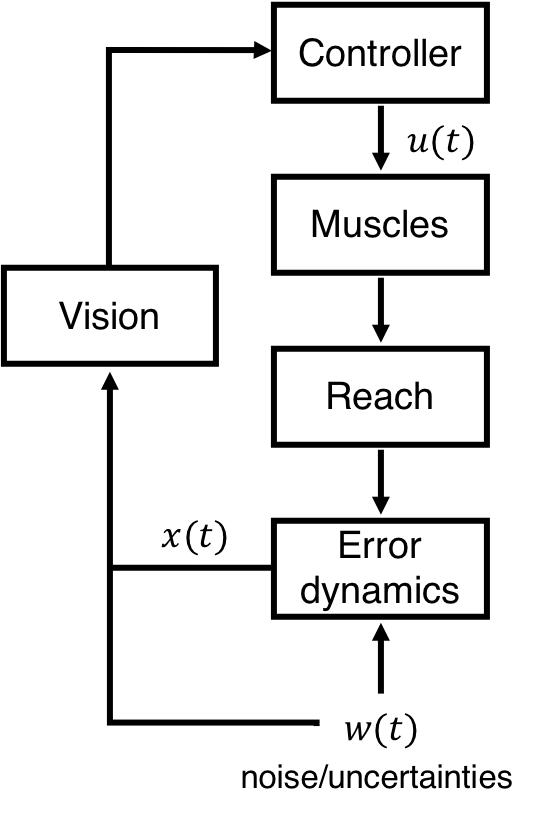}
\caption{Sensorimotor control model for reaching. Each box is a component that communicates (vision), computes (controller), or actuates (muscles) with potentially variable speed and accuracy. 
}
\label{figure:model}
\end{figure}

\begin{figure}
 \centering
\begin{flushleft}
\textbf{A}
\end{flushleft}
\center
\includegraphics[width=0.8\textwidth]{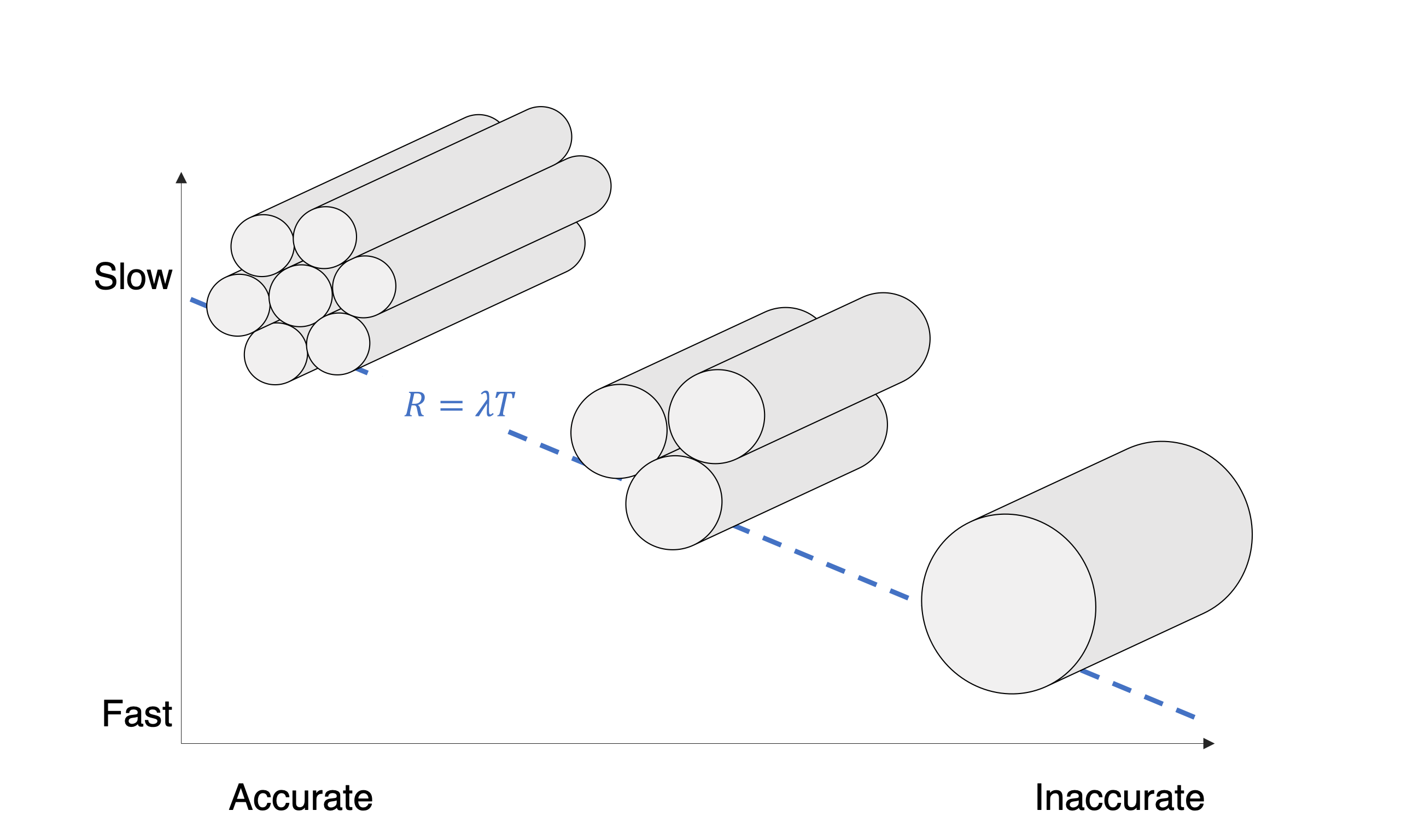}

\begin{flushleft}
\textbf{B}
\end{flushleft}
\center
\includegraphics[width=0.7\textwidth]{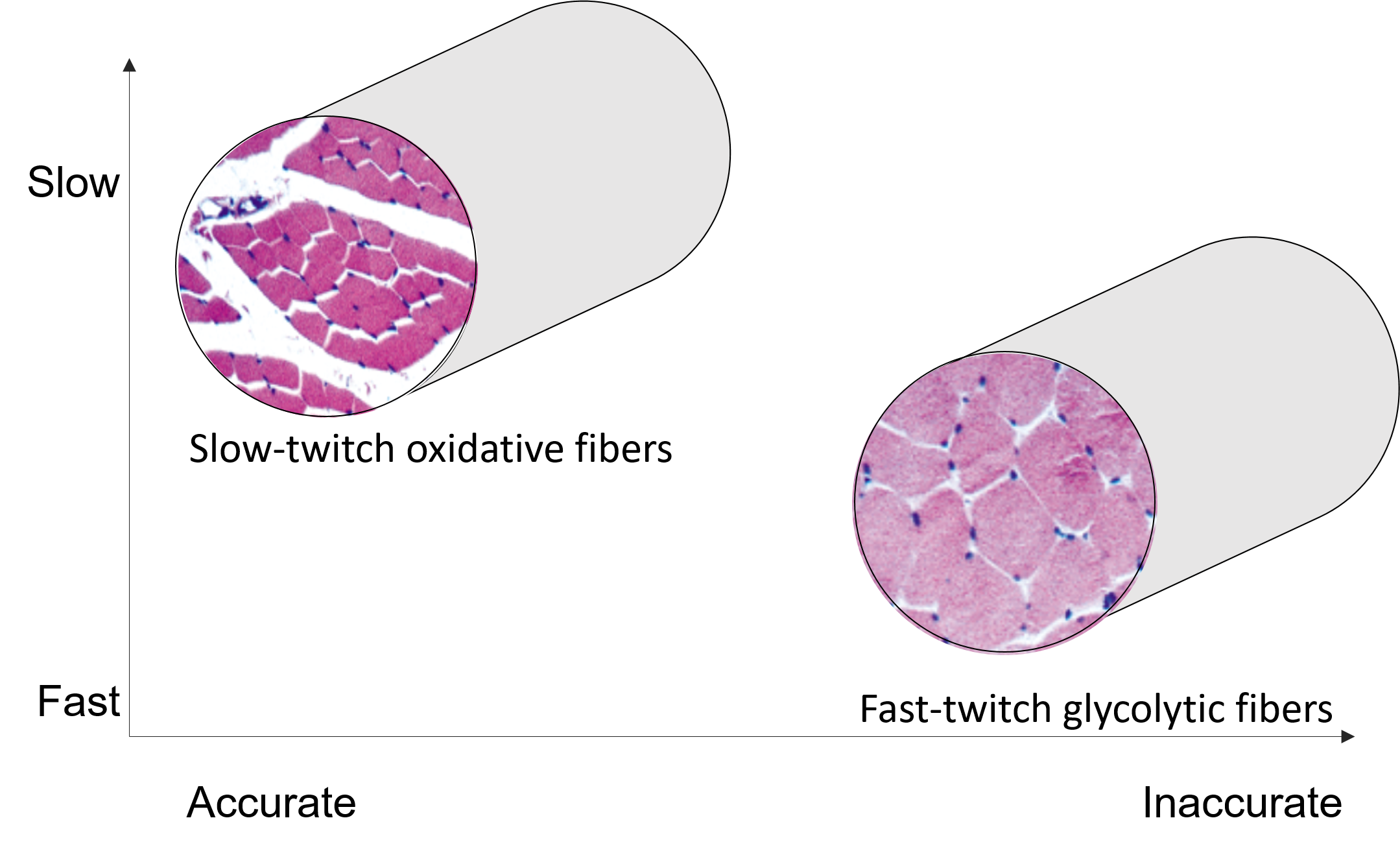}

 \caption{Component-level speed-accuracy trade-off (SAT) in nerves and muscles. The horizontal axis shows speed, and the vertical axis shows accuracy.
 (A) Signaling SATs of nerves with different axon size. The region above the dashed line represents the achievable speed and accuracy given a fixed total cross-sectional area, which is proportional to $\lambda$ in Eq.~\ref{eq:delay-rate-spike1}.
 (B) Actuation SATs of different muscle types. Muscles with smaller diameter and darker color (indicating larger amounts of myoglobin, mitochondria, and capillary density) contain oxidative fibers, whereas muscles with larger diameter and lighter color contain glycolytic fibers. Oxidative fibers are slower but more accurate than glycolytic fibers. Fast-twitch glycolytic fibers have two subtypes: fast oxidative glycolytic fibers (which use oxygen to help convert glycogen to ATP) and fast glycolytic fibers (which rely on ATP stored in the muscle cell to generate energy). 
}
 \label{figure:hardware-SAT}
\end{figure}

Consider the feedback control loop for reaching in Fig.~\ref{figure:model}. Here, the error between the actual position achieved in reaching and the desired position $x(t+1)$ is computed from the previous error $x(t)$, the sensed disturbance $w(t)$, and the control action $u(t)$ as follows:
\begin{equation}
\label{eq:plant}
x(t+1) =  x(t) + w(t) +  u(t) . 
\end{equation}
The control action, characterized by $\mathcal K$, is generated from the observed errors, sensed disturbance, and past control actions, \ie 
\begin{equation}
\label{eq:controller}
u(t+T) = \mathcal K( x(0:t),w(0:t-1) ,u(0:t+T -1)).
\end{equation}
using sensing components such as eyes and muscle sensors; communication components such as nerves; computing components such as the cortex in the central nervous system; and actuation components such as eye and arm muscles. The limitations of these components constrain the speed and accuracy of the feedback control loop. In particular, the feedback loop takes a total delay of $T := T_s + T_i$ for the new disturbance to be reflected in the control action, where $T_s$ captures the latency in nerve signaling, and $T_i$ captures other internal delays. Moreover, the feedback loop can only transmit $R$ bits of information per unit time (signaling rate). Table S1 in the Supplementary Material summarizes the above parameters.

The communication components--axons in sensory or motor neurons--carry sensory information from the periphery to the brain and activate muscles to execute the task. Heterogeneity in the size and number of axons within a nerve bundle and between different types of sensory nerves is characteristic anatomy, with calibers in mammals ranging over two orders of magnitude from tenths of microns to tens of microns~\cite{perge2012axons,stenum2018size,more2013sensorimotor,more2010scaling}. This size and number heterogeneity leads to extreme differences in neural signaling speed and accuracy, as the speed and rate of information flow in an axon depend on its diameter (and myelination). To quantify nerve signaling, we model axon bundles as a communication channel with signaling delay $T_s$ and signaling rate $R$. Building on~\cite{sterling2015principles}, we show in the Supplementary Materials that, under mild assumptions, the nerve signaling SAT can be modeled by 
\begin{equation}
\label{eq:delay-rate-spike1}
R = \lambda T_s  .
\end{equation}
where $ \lambda$ is proportional to the spatial and metabolic costs necessary to build and maintain the nerves. Nerve signaling SATs differ from species to species and increase with animal size~\cite{more2010scaling,big_animal}. Eq.~\ref{eq:delay-rate-spike1} can be refined or modified given specific types of nerves or encoding mechanisms, but the rest of our framework does not require the component SATs to have any specific form. For this reason, we will use Eq.~\ref{eq:delay-rate-spike1} to demonstrate how SATs at the component level impact those at the system level.

The actuation components -- muscles -- also display tradeoffs in terms of reaction speed, accuracy, maximum strength, and time course for fatigue. Striated muscles typically have both large fast-twitch fibers and smaller slow-twitch fibers (Fig.~\ref{figure:hardware-SAT}B). In particular, muscle SATs can be modeled using a simplified model that includes $m$ motor units, indexed by $i \in \{ 1 , 2, \cdots, m \}$, each associated with a reaction speed and a fixed strength. We use $F_i$ to denote its strength and assume without loss of generality that $F_1 \leq F_2 \leq  \cdots \leq F_m$. According to Henneman's size principle~\cite{henneman1965functional}, motor units in the spinal cord are recruited in ascending order of $F_i$, so a muscle (at non-transient time) can only generate $m+1$ discrete strength levels: $0$ and $\sum_{i = 1}^n F_i$ for $n = 1, 2, \cdots, m$. Given a fixed length, the maximum strength of a muscle $\ell = \sum_{i = 1 }^{m} F_i$ is proportional to its cross-sectional area~\cite{goldspink1985malleability}. This implies that, given a fixed space to build a muscle, its maximum strength does not depend on the specific composition of motor units. With cross sectional area constraining maximum strength, a muscle can be made of many motor units with small strengths or a few motor units with large strengths. In the former case, the muscle has better resolution (accuracy) but slower reaction speed, while in the latter case, the muscle is faster but performs at coarser resolution (see Fig. S2 in the Supplementary Material). This SAT can be quantified using the following formula: 
\begin{align}
\label{eq:muscle-a-pde}
&\dot a_i(t)  = \alpha f_i^p(t)  ( 1 - a_i(t) ) - \beta a_i(t) 
&a_i^q(t) = c_i(t) 
\end{align}
where $\alpha = 1, \beta = 1, p = 1, q = 3$ are fixed constants~\cite{brezina2000neuromuscular}. If a motor unit is recruited at time $t=0$, then its strength $c_i(t)$ rises according to Eq.~\ref{eq:muscle-a-pde} with $ f_i(t) = 1(t) /  (( 1 / F_i  )^{1/q} - 1) $, where $1(t)$ is a unit step function. Similarly, when a recruited motor unit is released at time $t= \tau$, its contraction rate falls according to Eq.~\ref{eq:muscle-a-pde} with $f_i (t)= 1(- t + \tau) /  (( 1 / F_i  )^{1/q} - 1)$. From Eq.~\ref{eq:muscle-a-pde}, the reaction speed of a muscle is an increasing function of $F_i$~\cite{note1f}, so better resolution (having small $F_i$) can only be achieved with decreased reaction speed.


Building on the basic model (Eq.~\ref{eq:plant} and Eq.~\ref{eq:controller}) we show how nerve SATs impact system SATs in reaching. In a reaching task, the subject's goal is to move their hand or cursor to a target as rapidly and accurately as possible. This goal can be achieved by setting $w(t) = d \delta (t)$ in Eq.~\ref{eq:plant}, where $d \in [-D , D]$ is the distance between the initial and target positions, and $\delta (t)$ is the Kronecker delta function~\cite{note4f}. There exist tradeoffs between reaching speed and accuracy, where the speed of reaching is quantified by the time taken to reach the target area $T_r$ and accuracy is quantified by the target width normalized by the distance $W/D$. The relation between $T_{r}$ and $D/W$ satisfies 
\begin{align} 
\label{eq:ent_time}
\sup_{|d| \leq D } T_{r} 
\geq T + \frac{1}{R} \log_2(2D/W) .
\end{align}
This formula explains Fitts' law~\cite{fitts1954information,wiki}, which states that the reaching time follows $T_r  = p + r \log_2(2D/W)$, where $F := \log_2(2D/W)$ is called the Fitts' index of difficulty, and $p$ and $r$ are fixed constants. The results in Eq.~\ref{eq:ent_time} suggest that the signaling delay $T_s$ affects reaching time $T_r$ in a linear manner, whereas the signaling rate $R$ affects $T_{r}$ in an inversely proportional manner. 
The proof of Eq.~\ref{eq:ent_time} can be found in the Supplementary Material. Intuitively, identifying a target of width $W$ in range $[-D, D]$ requires $F =  \log_2(2D/W)$ bits of information, and transmitting $F$ bits of information requires $F / R$ time steps with additional $T$ time steps of (transmission) delay in the feedback loop.

Eq.~\ref{eq:ent_time} decomposes into two terms: The term $T$ is a function only of delay and the term $\frac{1}{R} \log_2(2D/W)$ is a function only of the signaling rate. Therefore, we can consider the first term as the cost in reaching time due to delay in the feedback loop (denoted as the delay cost), and the second term as the cost due to limited signaling rate in the feedback loop (denoted as the rate cost). By combining the component-level SATs in Eq.~\ref{eq:delay-rate-spike1} and the system-level SATs in Eq.~\ref{eq:ent_time}, we can predict how the SATs in neural signaling impact sensorimotor control to obtain the worst-case reaching time as a function of the component speed and accuracy in Fig.~3A. Maximizing either the component speed or accuracy suffers from large delay or rate costs, which renders the system suboptimal. Thus, the reaching time is minimized when both the signaling delay and rate are 'tuned to' moderate levels.

In particular, subject to the nerve SAT Eq.~\ref{eq:delay-rate-spike1}, the reaching time is minimized at $T = \sqrt{F/\lambda}, R = \sqrt{\lambda F}$. The optimal speed is decreasing in $F$, whereas the optimal accuracy is increasing in $F$. This is because as the index of difficulty $F$ increases, the reaching task requires more accuracy, and the signaling rate limit gains greater impact on the reaching time than the signaling delay.

The dependencies of optimal nerve signaling speed and accuracy ($T, R$) on $F$ suggest that to have good reaching performance over a broad range of difficulties requires diversity in the signaling speed and accuracy. Such diversity in speed and accuracy can indeed be observed from the large heterogeneity in size and number of mammalian axons within a nerve bundle and between different types of sensory nerves (over two orders of magnitude)~\cite{stenum2018size,more2013sensorimotor,more2010scaling}. This size and number heterogeneity mediate wide ranges in neural signaling speed and accuracy because the speed and rate of information flow in an axon depend on its diameter (and myelination)~\cite{hodgkin1954note,hartline2007rapid}.



The analysis above shows the benefit of diversity in nerve signaling speed and accuracy on sensorimotor control, assuming that the SATs in nerve signaling are the bottleneck in the reaching task performance. Similar benefit of heterogeneous muscle speed and accuracy can also be observed in the control process when the performance bottleneck is muscle actuation.  
This point can be verified by the sensorimotor system model that is constrained by the muscle SATs.  

Specifically, we consider controlling the error dynamics Eq.~\ref{eq:plant} by the sensorimotor process Eq.~\ref{eq:controller} that are limited by the muscle actuation SATs Eq.~\ref{eq:muscle-a-pde}. We use this model to predict the reaching SATs when the muscle (actuation component) contains uniform motor units versus diverse motor units (Fig.~\ref{figure:DESS}B). Fig.~\ref{figure:DESS}B suggests that diversity of muscle components improves reaching performance over longer distances. Moreover, the resulting SAT resembles the logarithmic forms of Fitts' Law, which has a sweet spot that simultaneously attains both the speed and accuracy in reach. The sweet spot is achieved when large motor units facilitate fast activation at the beginning of the reaching, while small motor units are used to fine-tune the force toward the end of reaching. We term this \textit{diversity-enabled Sweet Spots} (DESS): The diversity in the component (muscle) speed and accuracy allows for fast and accuracy reaching.

We confirmed the benefit of diversity in motor units and muscles using reaching experiments (Figs.~3C,~3D,~S6).  This is done by using our open-source experiment platform, WheelCon~\cite{accexp}, whose setting and the visual interface is demonstrated in Fig.~\ref{figure:gaming-interface}. Subjects were asked to move to a target of fixed width as fast as possible. Fig.~\ref{figure:DESS}D compares reaching SATs for three different cases: When the subjects are allowed to use only a fast movement speed, only a slow speed, and both. The reaching times with fast and slow speeds were generally shorter than those either using fast or slow speed for all settings of Fitts' difficulty $(D/W)$, suggesting that diversity enables a sweet spot. Moreover, the reaching time is maintained around similar levels for all levels of $D/W$ when diverse speeds are allowed, whereas the reaching time does change for different levels of $D/W$ when only uniform movement speed is allowed. This phenomenon indicates that diversity is key for the Fitts' law like performance: The reaching time stays the same under varying $D, W$ pairs with fixed $D/W$. The utilization of speed and accuracy diversity can indeed be observed in the reaching experiment (Fig.~S5): Subjects typically use fast speed in the beginning and then slow down to achieve better accuracy toward the end.  
These results together suggest that Fitts' Law describes a diversity-enabled sweet spot in sensorimotor control.

We evaluated the theoretical prediction Eq.~\ref{eq:ent_time} with reaching experiments that tested the sensorimotor control loop in Fig.~\ref{figure:DESS}A. Testing the impact of hardware speed and accuracy requires us to measure the reaching time under different delay and inaccuracies of the sensorimotor feedback control loop. As we cannot change the internal speed and accuracy of each subject's sensorimotor control system, instead we add external delays and inaccuracies to change the speed and accuracy of the sensorimotor feedback control loop (see the Supplementary Material for details).  
Fig.~\ref{figure:DESS} C shows the reaching times when subjects use a steering wheel to move a cursor from an initial position to a target. External delay and/or quantization were artificially inserted between the wheel and cursor movement (see the Supplementary Material for details) to create a specific SAT that can be manipulated in the task. The result in Fig.~\ref{figure:DESS}C suggests that subjects can adapt to the external SATs and perform as the theory predicts in Eq.~\ref{eq:ent_time}. 

This benefit of diversity can also be found at the level of the whole arm, which combines large arm muscles and joints with small fingers and fine articulation. Further, a combination of fast coarse movements and slow accurate ones is known to produce a logarithmic law~\cite{meyer1988optimality}, which has a sweet spot that achieves both speed and accuracy. DESSs theory may help us understand how engineered systems can be designed to achieve fast and accurate performance with slow or inaccurate components. 
Take the long-distance travel as an example. No single mode of transportation (\textit{e.g.} walking, driving, flying) can rapidly and accurately transport you to your destination. But a combination of flying rapidly between cities, driving within a city, and walking to the destination can together achieve a DESS of efficient, accurate transport (see the Supplementary Material for more detail). 

Our results suggest how Fitts' Law could arise from DESSs, in which hardware \textit{diversity} is key for achieving fast and accurate performance using slow or inaccurate hardware. This relationship between DESSs and logarithmic laws could provide new insights into other systems with logarithmic laws. Examples of logarithmic laws are the Weber-Fechner Law for the relation between the physical change in a stimulus and the perceived change in human perception;  the Ricco Law for visual target detection for unresolved targets; the Accot-Zhai Law for steering (a generalization of Fitts' Law for 2D environments); the spacing effect of Ebbinghaus for long-term recall from memory; and the Hick-Hyman law for the logarithmic increase in the time it takes to make a decision as  the number of choices increases (see~\cite{fitts1954information,kvaaiseth1979note,mackenzie1989note,hatze1979teleological,wiki} and references therein).

Although this paper focuses on the benefit of hardware diversity and its connection to Fitts' Law for reaching, DESSs are widely observed in the layered architectures which use different types of control, such as the control of eye movements~\cite{lac1995learning,lisberger2010visual}, decision-making~\cite{kahneman2011thinking,franks2003speed,chittka2009speed}, and many other behaviors with speed-accuracy tradeoffs (see~\cite{heitz2014speed} and references therein). For example, the oculomotor system has multiple layers: The vestibulo-ocular reflex layer performs fast but inaccurate negative feedback control to stabilize images on the retina despite rapid head movements. This layer works in concert with another layer that performs smooth pursuit, a slow but accurate cortical system for visual tracking of slowly moving objects. These two layers jointly create a virtual eye controller that is both fast and accurate. More detail about DESSs in layered control architectures like the visual system is presented in our companion paper~\cite{pnas}. Importantly, DESSs reveal a general design principle for distributed control in brains that can inspire the design of high-performance, large-scale technological systems.

\newpage

\bibliography{scibib}
\bibliographystyle{Science}

\section*{Acknowledgments}
This research was supported by National Science Foundation (NCS-FO 1735004 and 1735003) and the Swartz Foundation. Q.L. was supported by a Boswell fellowship and a FWO postdoctoral fellowship (12P6719N LV).



\clearpage
\section*{Figure Legends}
\setlength{\parindent}{0em}
\textbf{Figure 1}: Sensorimotor control model for reaching. Each box is a component that communicates (vision), computes (controller), or actuates (muscles) with potentially variable speed and accuracy. 

\textbf{Figure 2}: Component-level speed-accuracy trade-off (SAT) in nerves and muscles. The horizontal axis shows speed, and the vertical axis shows accuracy.
 (A) Signaling SATs of nerves with different axon size. The region above the dashed line represents the achievable speed and accuracy given a fixed total cross-sectional area, which is proportional to $\lambda$ in Eq.~\ref{eq:delay-rate-spike1}.
 (B) Actuation SATs of different muscle types. Muscles with smaller diameter and darker color (indicating larger amounts of myoglobin, mitochondria, and capillary density) contain oxidative fibers, whereas muscles with larger diameter and lighter color contain glycolytic fibers. Oxidative fibers are slower but more accurate than glycolytic fibers. Fast-twitch glycolytic fibers have two subtypes: fast oxidative glycolytic fibers (which use oxygen to help convert glycogen to ATP) and fast glycolytic fibers (which rely on ATP stored in the muscle cell to generate energy). 

\textbf{Figure 3}: Theoretical and empirical reaching SATs and DESSs. (A) Theoretical SATs in the reaching task. The delay cost (blue line), rate cost (red line), and total cost (black line) in Eq.~\ref{eq:ent_time} are shown with a varying delay and data rate, which satisfy the component SAT $T  = (R-1)/8$. 
(B) Theoretical DESSs in the reaching task. The SAT between normalized width and reaching time is obtained from a sensorimotor control model involving a muscle with uniform motor units (dashed line) or diverse motor units (solid line). The DESSs for two feedback loops with diverse muscles and uniform muscles are shown in Fig S7. (C) Empirical SATs in the reaching task. Data obtained from six subjects who performed the task over a range of time delays and levels of quantization (See Fig.~S1 for data from individual subjects). The performance with added actuation delay $T$ is shown in the solid blue line; the performance with added quantizer of rate $R$ in shown in the solid red line. The performance with added delay and quantization satisfying $T = (R-1)/8$ is shown in the solid black line. The standard errors are shown in the shaded region around the solid lines. 
(D) Benefit of using diverse speeds in the reaching task. The plot shows the performance of four subjects who performed the reaching task with uniformly fast speed (orange line), uniformly slow speed (blue line), and both fast and slow speeds (green line). The standard errors across subjects are shown in the shaded region around the solid lines. 

\textbf{Figure 4}: Illustration of the reaching experiment performed. (A) The set-up: a subject watching a monitor and steering a wheel. (B) Video interface for the reaching experiment. The green line indicates the player's position and the gray zone indicates the target. The subject's goal is to reach the target as fast as possible and stay at the target by controlling the steering wheel. (C) Error dynamics in three conditions. The error (denoted by $x(t)$ in the text) is defined to be the difference between the player's position and the center of the target. The solid line and the shadow indicate the averaged error trajectory and the standard errors for the trials with $D=4$ and $W=1$, respectively.

\clearpage

\begin{figure}
 \centering

\begin{flushleft}
  \textbf{A}
\end{flushleft}
\center
\includegraphics[width=0.5\textwidth]{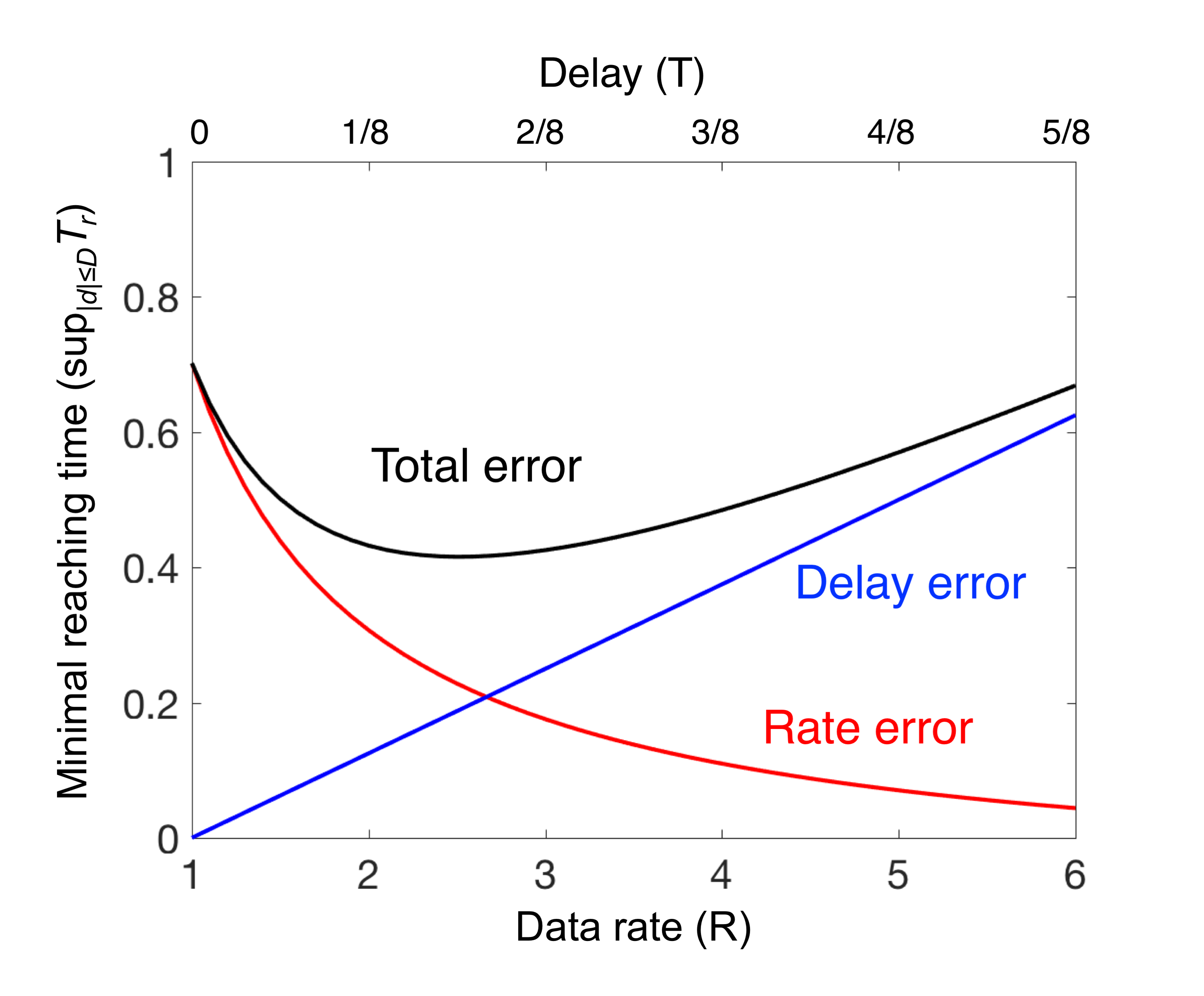}

\begin{flushleft}
  \textbf{B}
\end{flushleft}
\center
\includegraphics[width=0.5\textwidth]{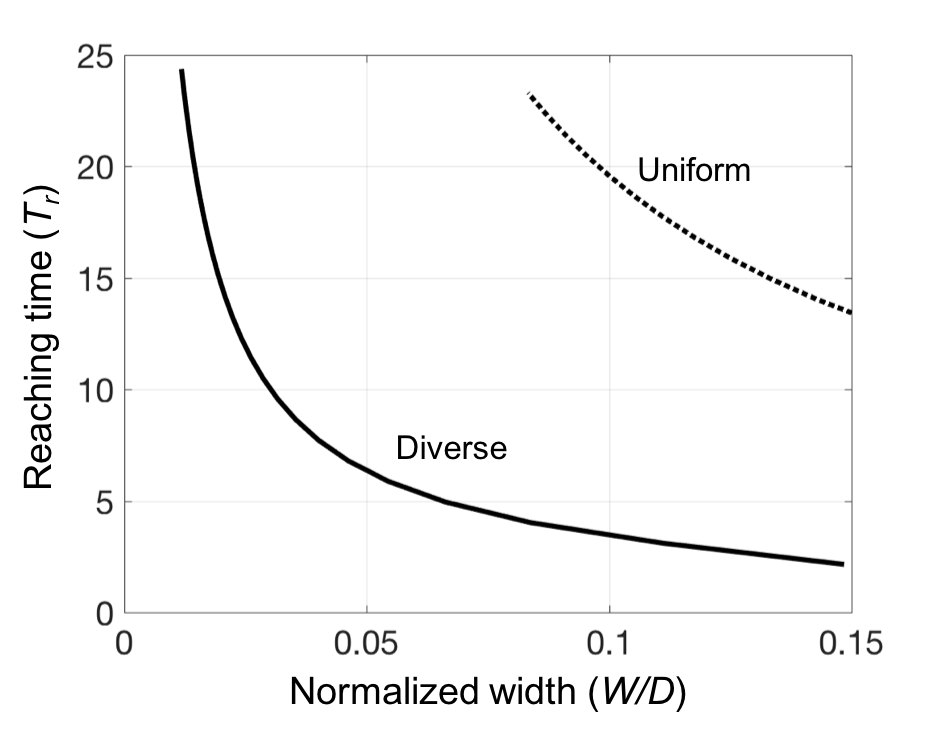}
\end{figure}

\begin{figure}
\begin{flushleft}
  \textbf{C}
\end{flushleft}
\center
\includegraphics[width=0.5\textwidth]{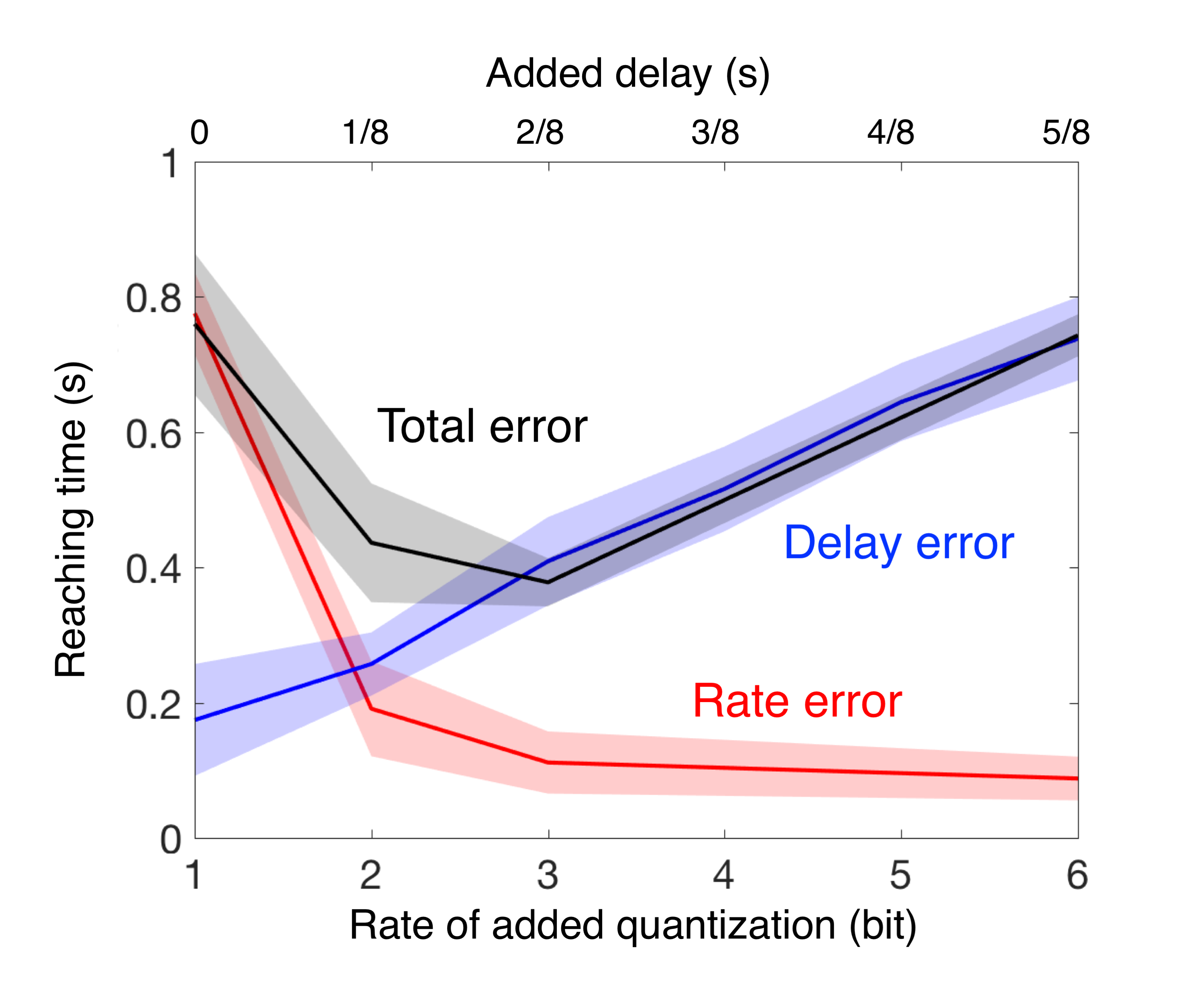}

\begin{flushleft}
  \textbf{D}
\end{flushleft}
\center
\includegraphics[width=0.5\textwidth]{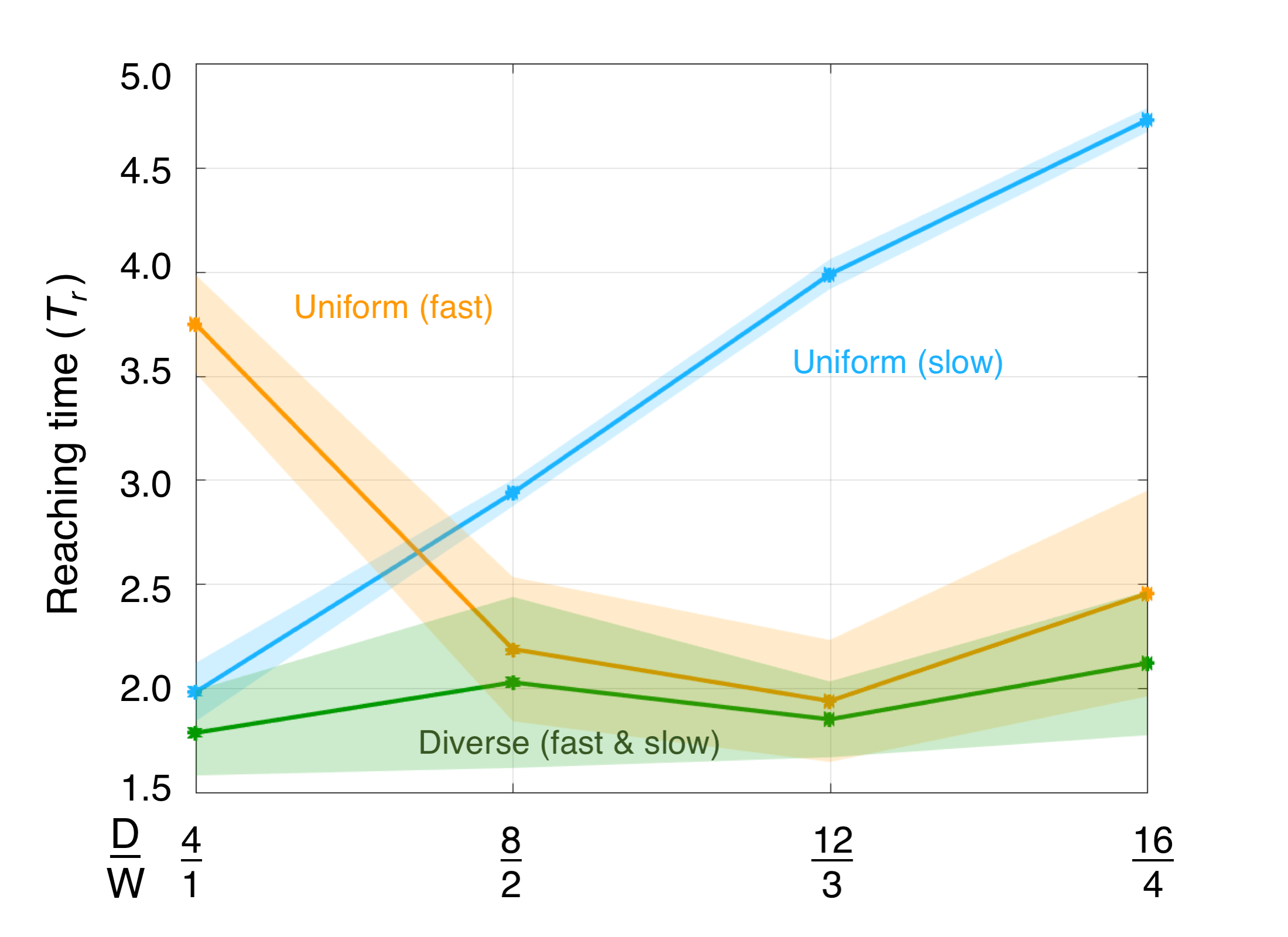}

\caption{
Theoretical and empirical reaching SATs and DESSs. (A) Theoretical SATs in the reaching task. The delay cost (blue line), rate cost (red line), and total cost (black line) in Eq.~\ref{eq:ent_time} are shown with a varying delay and data rate, which satisfy the component SAT $T  = (R-1)/8$. 
(B) Theoretical DESSs in the reaching task. The SAT between normalized width and reaching time is obtained from a sensorimotor control model involving a muscle with uniform motor units (dashed line) or diverse motor units (solid line). The DESSs for two feedback loops with diverse muscles and uniform muscles are shown in Fig S7. (C) Empirical SATs in the reaching task. Data obtained from six subjects who performed the task over a range of time delays and levels of quantization (See Fig.~S1 for data from individual subjects). The performance with added actuation delay $T$ is shown in the solid blue line; the performance with added quantizer of rate $R$ in shown in the solid red line. The performance with added delay and quantization satisfying $T = (R-1)/8$ is shown in the solid black line. The standard errors are shown in the shaded region around the solid lines. 
(D) Benefit of using diverse speeds in the reaching task. The plot shows the performance of four subjects who performed the reaching task with uniformly fast speed (orange line), uniformly slow speed (blue line), and both fast and slow speeds (green line). The standard errors across subjects are shown in the shaded region around the solid lines. }
 \label{figure:DESS}
\end{figure}

\begin{figure}
\centering
\begin{flushleft}
  \textbf{A}
\end{flushleft}
 \center
\includegraphics[width=0.5\textwidth]{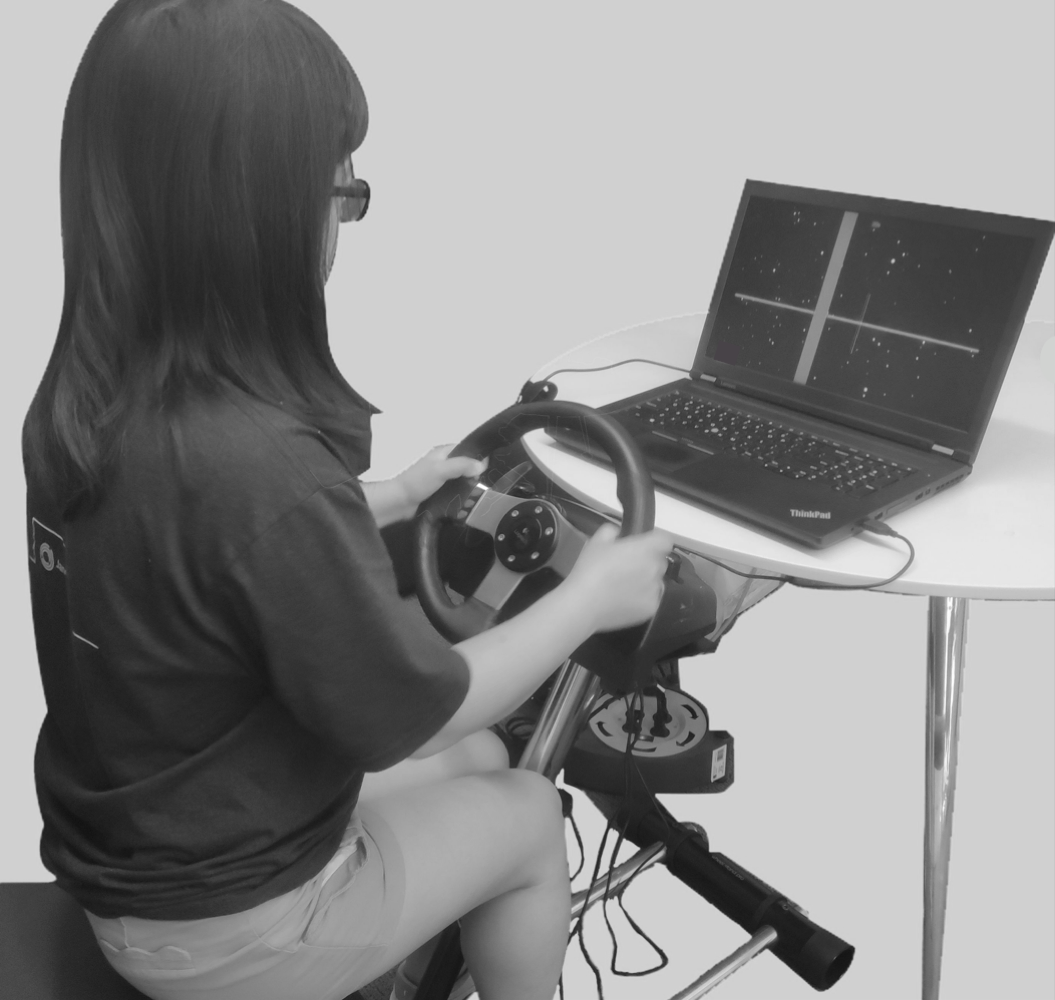}
 
\begin{flushleft}
\textbf{B}
\end{flushleft}
 \center
\includegraphics[width=0.5\textwidth]{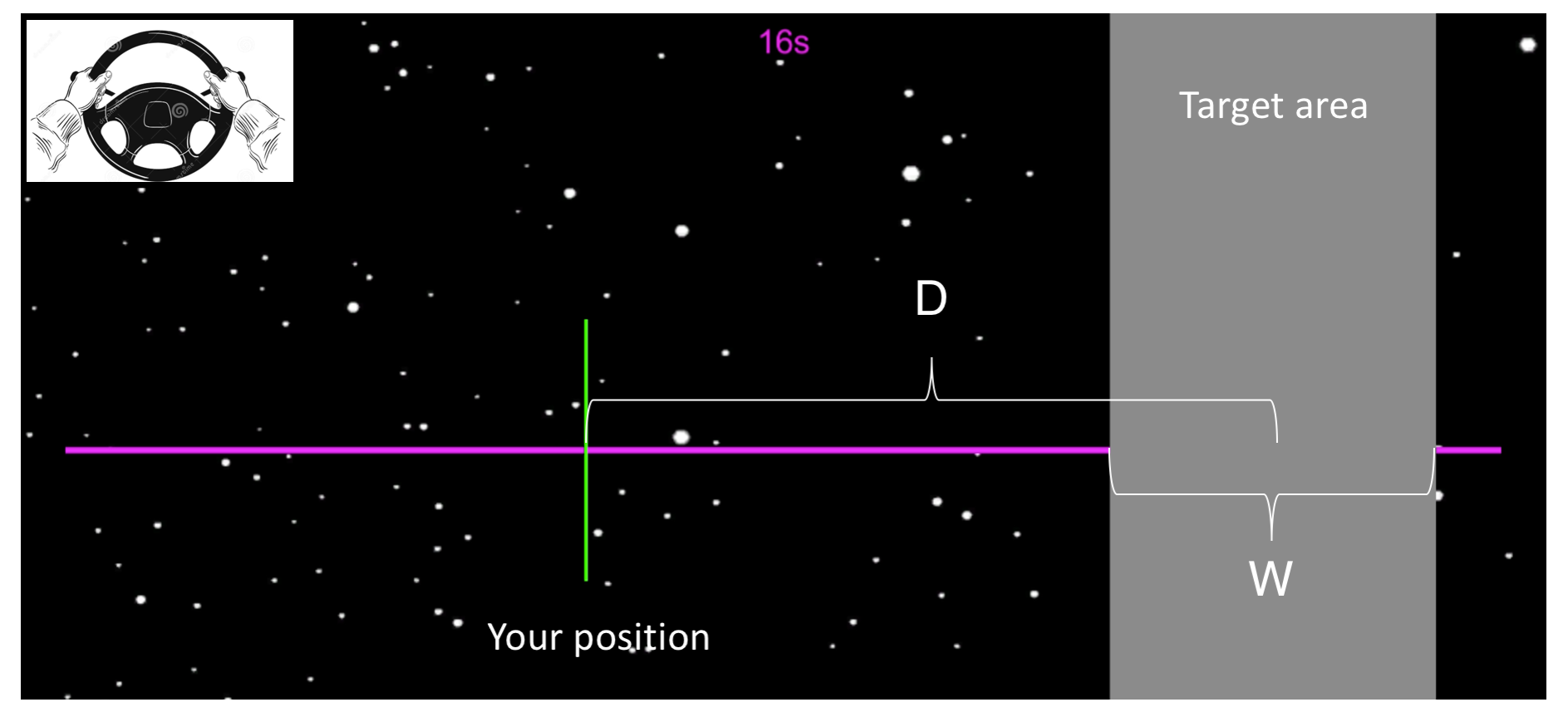}
 \end{figure}

\begin{figure}
\begin{flushleft}
  \textbf{C}
\end{flushleft}
\center
\includegraphics[width=0.5\textwidth]{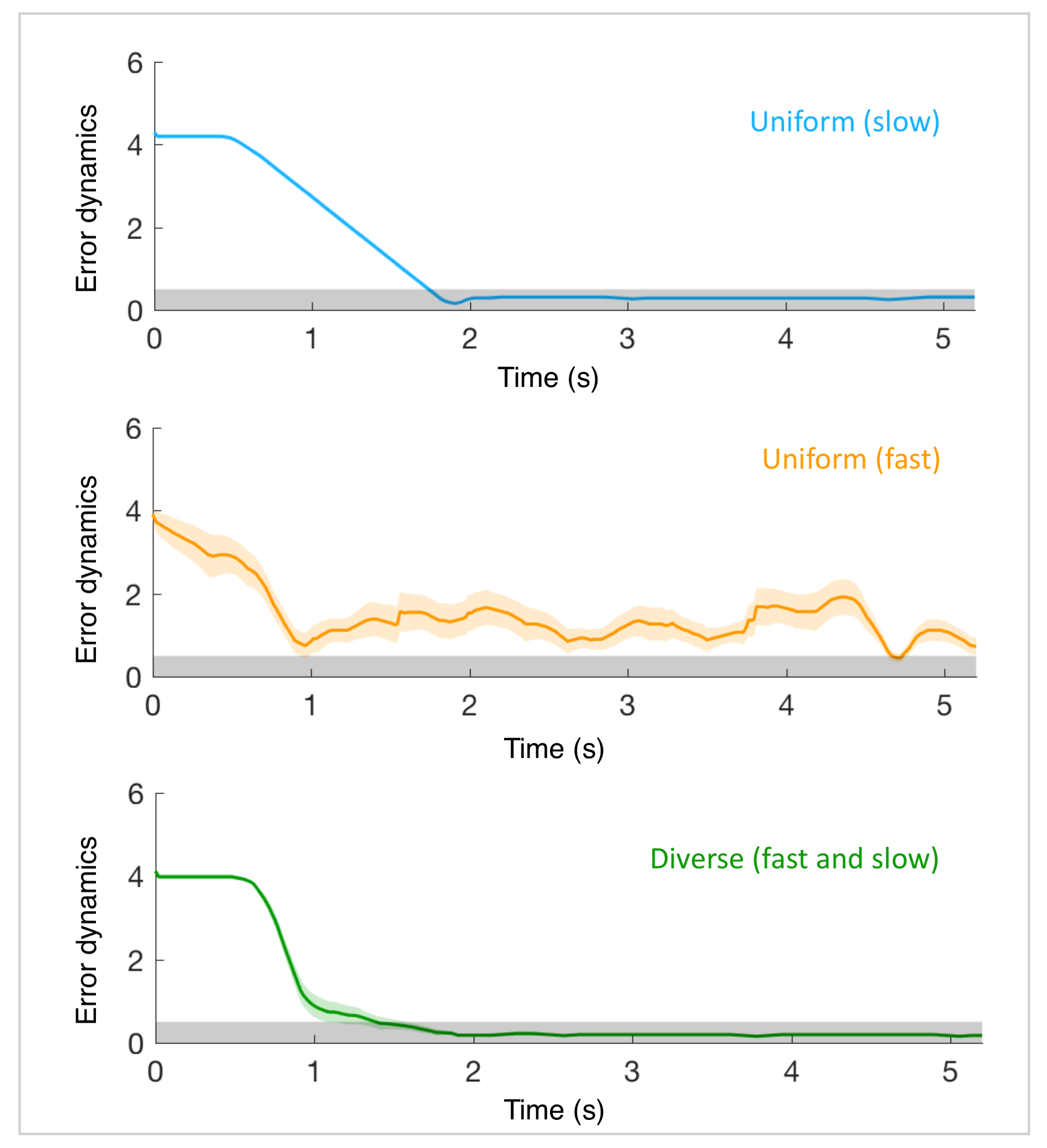}
 
\caption{Illustration of the reaching experiment performed. (A) The set-up: a subject watching a monitor and steering a wheel. (B) Video interface for the reaching experiment. The green line indicates the player's position and the gray zone indicates the target. The subject's goal is to reach the target as fast as possible and stay at the target by controlling the steering wheel. (C) Error dynamics in three conditions. The error (denoted by $x(t)$ in the text) is defined to be the difference between the player's position and the center of the target. The solid line and the shadow indicate the averaged error trajectory and the standard errors for the trials with $D=4$ and $W=1$, respectively. 
}
 \label{figure:gaming-interface}
\end{figure}


\clearpage
\section*{Supplementary Material}
The nerve signaling SAT\\
The impact of nerve SATs on reaching SATs\\
The impact of muscle SATs on reaching SATs\\
Experiments \\
Diversity-enabled sweet spots in transportation\\
Figs. S1-S8\\
Table S1\\
Video S1\\
References and notes

\setcounter{table}{0}
\makeatletter 
\renewcommand{\thetable}{S\@arabic\c@table}
\makeatother

\begin{table}
\begin{center}
\begin{tabular}{@{\extracolsep{5pt}}ll} 
\\[-1.8ex]\hline 
\hline \\[-1.8ex] 
\textbf{Parameter} & \textbf{Description} \\
\hline 
$x(t)$ & Error at time step $t$  \\
\hline
$\mathcal K$ & Controller  \\
\hline
$T_s \geq 0$ & Signaling delay  \\
\hline
$T_i \geq 0$ & Internal delay \\
\hline
$T = T_s + T_i$ & Total delay 
\\
\hline
$T_t$ & Time to reach target \\
\hline
$R$ & Information rate (bits per unit time) 
\\
\hline
$\lambda$ & Cost associated with the resource use
\\
\hline
\end{tabular}
\caption{Parameters in the basic model.
}
\label{tab:notations}
\end{center}
\end{table}

\bibliography{scibib}
\bibliographystyle{Science}

\clearpage

\section*{Supplementary Materials}

\subsection*{The nerve signaling SAT (Eq. 3 in the main text)}

In this section, we characterize the SATs for neural signaling. The formula for the SATs depend on how the nerves encode information (\eg spike-based, spike-rate, etc.). Here, we focus on spike-based encoding and discuss alternative encoding strategies in our companion paper \cite{pnas}. In a spike-based encoding scheme, information is encoded in the presence or absence of a spike in specific time intervals, analogous to digital packet-switching networks~\cite{salinas2001correlated, srivastava2017motor}. This encoding method requires spikes to be generated with sufficient accuracy in timing, which has been experimentally verified in multiple types of neuron  \cite{mainen1995reliability,fox2010encoding}. We consider a nerve with bundles of axons. We use $n$ to denote the number of axons in a nerve and $\rho$ to denote their average radius. We use $T_s, R$ to denote the delay and data rate (\ie the amount of information in bits that can be transmitted) of a nerve. When the signaling is precise and noiseless, an axon with achievable firing rate $\phi$ can transmit $\phi$ bits of information per unit time. For sufficiently large myelinated axons, we assume that the propagation speed $1/T_s$ is proportional to the axon radius $\rho$~\cite{sterling2015principles}, \ie 
\begin{align}
\label{eq:alpha-beta}
&T_s = \alpha/\rho
\end{align}
for some proportionality constant $\alpha$. We also model the achievable firing rate $\phi$ to be proportional to the axon radius $\rho$, \ie 
\begin{align}
&\phi=\beta \rho,
\end{align}
for some proportionality constant $\beta$. Moreover, the space and metabolic costs of a nerve are proportional to its volume~\cite{sterling2015principles}, and given a fixed nerve length, these costs are also proportional to its total cross-sectional area $s$. Using the above properties, we have 
\begin{align}
R = n \phi = \frac{s}{\pi \rho^2} \beta \rho = \frac{s \beta}{\pi} \frac{1}{\rho} =  \frac{s \beta}{\alpha \pi} T_s. 
\end{align}
This leads to $R = \lambda T_s$, where $ \lambda  = s \beta / \pi \alpha $ is proportional to the spatial and metabolic costs to build and maintain the nerves.


\subsection*{The impact of nerve SATs on reaching SATs}

We propose a control system model in the deterministic worst-case setting that captures the nerve SATs. Toward this end, we consider a reaching task in which the subject needs to move to a given target as quickly as possible. The target is chosen from a set of disjoint intervals of length $W$ in a range of distance $D$ from the origin, \ie $[-D, D]$. We define the reaching time as 
\begin{align}
T_r = \{ \tau : |x(t)| \leq W/2 \text{ for any } t \geq \tau, |x(0)| \leq D \}.  
\end{align}
The range $[-D, D]$ can hold no less than $2D / W$ disjoint intervals of length $W$, so the amount of information required to differentiate one interval from other such intervals can be computed to be $F = \log_2 ( 2D / W )$ bits. This means that, after the target location is sensed, $F / R$ time steps are required to deliver $F$ bits of information from the sensors to the actuators. Then, there is a delay of $T$ time steps before it is reflected in the actuation.  

Therefore, the worst case reaching time cannot be smaller than the sum of the two types of delays, $T + F/R$, yielding Eq. 5 in the main text. For a more general setting, the applicable math tools can be found in our companion paper \cite{pnas}.  

As suggested by the existing literature, in the setting described in this paper, $F$ quantifies the amount of information required to perform the reaching task in the required accuracy. See \cite{fitts1954information,mackenzie1989note,kvaaiseth1979note,wiki,goldberg2015two} and references therein for more details on how in stochastic settings $F$ and related quantities are associated with the required information. Moreover, the above proof provide insights into Fitts' law, \ie $T_r = p + q F$, and how the nerve signaling SATs impact the reaching SATs: $ p = T$ is determined from the delay in reacting and transmitting the target information, and $q = 1/R$ is determined from the limited data rate in the feedback loop.

\subsection*{The impact of muscle SATs on reaching SATs}

Existing literature has studied Fitts' law using conventional control theory that does not account for explicit hardware constraints. For example, conventional control theory can model jerk~\cite{crossman1983feedback}, smoothness~\cite{crossman1983feedback}, acceleration~\cite{flash1985coordination}, and kinematics~\cite{plamondon1997speed}, among others. Although quantities like jerk may be reduced as a result of optimizing control performance subject to hardware constraints, they are unlikely to be the ultimate target of optimization. In order to connect the hardware and reaching SATs, we propose a control system model that captures the muscle SATs explicitly. Toward this end, we use a simplified muscle model that includes $m$ motor units. Each motor unit is indexed by $i \in \{ 1 , 2, \cdots, m \}$ and is associated with a reaction speed and a strength level. We use $F_i$ to denote its strength and assume without loss of generality that $F_1 \leq F_2 \leq  \cdots \leq F_m$. Recall from the main text that the muscle SAT can be quantified using the following formula:
\begin{equation}
\label{eq:muscle-a-pde}
\begin{aligned}
& \frac{ d }{dt} a_i(t)  = \alpha f_i^p  ( 1 - a_i(t) ) - \beta a_i(t) \\
&a_i^q(t) = c_i(t) 
\end{aligned}
\end{equation}
where $\alpha = 1, \beta = 1, p = 1, q = 3$, and $f_i = 1/  (( 1 / F_i  )^{1/q} - 1)$ are fixed constants~\cite{brezina2000neuromuscular}. See Fig. \ref{fig:muscle-SAT} for an illustration of the dynamics of Eq. \ref{eq:muscle-a-pde}. If a motor unit is recruited at time $t=0$, then its strength $c_i(t)$ rises according to Eq. \ref{eq:muscle-a-pde} and $c_i(t) = 1(t) F_i$ as follows: 
\begin{equation}
\label{eq:muscle-b-pde}
\begin{aligned}
 c_i(t) = \left\{  \frac{f}{f+1} ( - e^{-t(f+1)} + 1) \right\}^{1/q} 
\end{aligned}
\end{equation}
where $1(t)$ is a unit step function. Similarly, when a recruited motor unit is released at time $t=0$, its contraction rate falls according to Eq. \ref{eq:muscle-a-pde} with $f_i(t) = 1(- t) /  (( 1 / F_i  )^{1/q} - 1)$ as follows:
\begin{align}
 c_i(t) = c_i(0) e^{-t} .
\end{align}
Next, we consider using the muscle to perform a reaching task. The control process in reaching is modeled by Eq 1-2 in the main text. Recall that $x(t)$ denotes the error between the actual position and the target (desired) position. Given a static target, the error dynamics only depends on the actual position. So, we have 
\begin{align}
\label{eq:error-dynamics-motor}
\frac{ d^2  }{ dt^2 } x(t) = \sum_i c_i(t) - h(t)
\end{align}
where the sum is taken over all recruited motor units. The function $h(t)$, which captures the friction acting against the motion, takes the form 
\begin{align}
 h(t) = \begin{cases}
    h_s & \text{if } d x(t) / dt = 0\\
    h_k & \text{otherwise}   \end{cases},
\end{align}
 where $h_s$ can be obtained from the coefficient of static friction, and $h_k$ can be obtained from the kinetic friction.

To study the benefit of diversity in motor units, we consider two cases: having two mid-sized motor units (denoted as the uniform case), having a large motor unit and a small one (denoted as the diverse case). In both cases, the total resource use (total cross-sectional area of all motor units) is set to be equivalent, and so does the sum of strength levels for both cases. Specificaly, the uniform muscle case has two motor units with the same strength level ($F_1=F_2=0.5$), and the diverse case has two motor units with different strength levels ($F_1=0.85, F_2=0.15$). Fig.~S3 (A) shows the achievable reaching time and distance as dots when each motor unit contracts for periods whose length ranges from $0.75$ to $14.75$ with $0.5$ increment. This increment in duration captures the latency and limitations on timing precision in reaction. In the diverse case, Fig.~S3B zooms in the plots in Fig.~S3A and compares the achievable reaching times between the uniform cases and diverse cases when the contraction duration is set to be equal. From Fig.~S3, we can observe that having diverse motor units is beneficial. Compared with the uniform case, large motor units in the diverse case are helpful for reaching longer distance due to fast activation, while the small motor units can be used to achieve precise movements. This benefit of reaching SATs in diverse motor units is shown in Fig.~3A of the main text. 

To study the benefit of diversity in muscles, we consider two cases: Having two muscles with the same strength level (denote as the uniform case), having two muscles with different strength levels (denote as the diverse case). The total strength for each case is set to be same, so the total cross-sectional area used by the muscles in each case is also equivalent. Fig. S4 compares the reaching SATs for each case. Similarly, having diverse muscles is beneficial because the muscle with larger strength in the diverse case is helpful for achieving faster activation, and the muscle with smaller strength is helpful for achieving precise movements. Although existing literature has studied the relation between motor variability and Fitts-like laws ~\cite{faisal2008noise,schmidt1979motor,van1995fitts}, our results provide the new insights that diversity in muscle or nerve is essential in obtaining Fitts-like laws.

\subsection*{Experiments} 
We designed several reaching tasks to validate our theory for the sensorimotor SATs and DESS. All experiments involved in this study have institutional review board (IRB) approval from California Institute of Technology. All participants signed the informed consent before participating tasks. Testing the effects of hardware speed and accuracy on sensorimotor performance is generally hard because we cannot vary the internal delay and data rate in real nerves or muscles. Alternatively, we externally add delay, quantization, speed, or resolution constraints in vision or motion. For sufficiently large external constraints, we can vary such external constraints to study their impact on sensorimotor performance. 

First, we tested the nerve signaling SATs on sensorimotor performance. Subjects are asked to perform the task of steering a wheel to reach (and stay) in the target gray zone as quickly as possible. To test the effect of having a delay and limited data rate in the feedback loop, we conducted two types of experiments: (1) reaching with added delay, (2) reaching with added quantization, (3) reaching with added delay and quantization with the SAT constraint $T=(R-1)/8$.  In case (1) , the visual display was delayed for $0, 1/8, \cdots, 5/8$ seconds. 
In case (2), the wheel input is quantized by $1, 2, \cdots, 6$ bits per unit sampling interval, where the sampling interval is set to be $350ms$. Specifically, a quantizer of data rate $R$ is implemented as follows: after $n$ sampling intervals, the gray zone is centered at the target position with a width of the screen length times $2^{-n R}$. This intends to simulate the process of sending $R$ bits per sampling interval and estimating the target with an error of size $2^{-n R}$ after $n$ sampling intervals (see Supplementary video S1). 
In case (3), both the external delay and quantization was added, where the external delay $T$ and quantizer rate $R$ are set to satisfy the SAT $T=(R-1)/8$. Each case (1)--(3) is tested for 50 trials per subject, and their reaching times are recorded. The subject-specific internal delay was estimated to have mean = $1.17$s, SD = $0.06$ for each subject by the minimal time to reach the target area with no additional delay or quantization. The internal delay was subtracted for the following analysis. Plots of the mean movement time from single subjects are shown in Fig. S1, and the average of all six subjects is shown in Fig.~3C in the main text. The blue, red, and black lines in each plot shows the result from case (1)--(3) respectively. 

Second, we tested the benefit of muscle diversity in reaching tasks for two settings: by varying the target distance and width, and by varying the target width for fixed distance. In the first setting, we varied the target distance with $D=4,8,12,16$ and the corresponding target width with $W=1,2,3,4$, which results in a fixed normalized width $W/D$. We considered three cases: (1) uniformly slow speed ($|V_\text{slow}|=2.5$), uniformly fast speed ($|V_\text{fast}|=10$), and (3) diverse speeds ($|V_\text{slow}|=2.5$ and $|V_\text{fast}|=10$). The speed is controlled by the subjects through the wheel angle as in in Fig.~S5. Here, the unit of the speed is a screen unit per sampling intervals, where a screen unit is set so that the total size of $1000$ units sum up to equal the whole screen monitor ($15$ inches with $1920 \times 1080$ resolution), and the sampling interval is set to be $0.01$ seconds. The performance from both cases is shown in Fig.~3D. The diverse case performs much better than the uniform case. This is because subjects in the uniform case can only use the lower speed to satisfy the required accuracy, whereas subjects in the diverse case can achieve better reaching SATs by starting with the higher speed to reach the target quickly and then switching to the lower speed to stay inside the desired zone (Fig.~S6). The performance at single subject level is shown in Fig.~S7.

In the second setting, we set the target distance to be $D=12$ and varied the target width to be $W=1,2,3,4$. We considered two cases: (1) when the subject only has one choice of speed ($|V_0|=2.5$), and (2) when the subject can choose from two speeds ($|V_0|=2.5$ and $|V_1|=5$). Specifically, in the uniform case, the speed is set to be $V_0=-2.5$ when the subject steers to the left side (angle from the middle $\leq 0^{\circ}$), and $V_0=2.5$ when the subject steers to the right side (angle from the middle $>0^{\circ}$). In the diverse case, the speed is controlled by the wheel angle as follows
\begin{align}
V =  
\begin{cases}
-5 & \text{angle} \leq -30 \\
- 2.5 & -30 \leq  \text{angle} \leq 0 \\
2.5 & 0 \leq \text{angle} \leq 30 \\
5 & 30 \leq \text{angle} 
\end{cases} .
\end{align}
The performance from both cases is shown in Fig.~S8B.

\subsection*{Diversity enabled sweet spots in transportation}

DESS can be observed in transportation as well. To see this, let's consider a simple transportation model of traveling by walking, driving, or flying. We index these means of transportation by $i = 1,2,3$ respectively and use $s_i$ to denote their speed and $e_i$ to denote their resolution. As walking is typically slower than driving, and driving is slower and flying, we assume that $s_1 < s_2 < s_3$. Meanwhile, as flights can only land on airports, cars can only stop at parking lots, and a walker can stop at almost anywhere, we assume that $e_1 < e_2 < e_3$. Let $T_E(D)$ be the time to travel distance $D$ with tolerable error $E$, where $E$ satisfies $E \ll D$ and $E\geq r_1$. When the traveler is only allowed to use a single means of transportation, the relation between the traveling time $T_E(D)$ and accuracy constraint $E$ follows
\begin{equation}
 T_E (D) = 
  \begin{dcases}
D/s_1  + O(1)       & \text{if }  e_1 \leq E \leq e_2 \\
D/s_2  + O(1)         & \text{if } e_2 \leq E \leq e_3 \\
D/s_3 + O(1)        & \text{if } e_3 \leq E ,  \\
  \end{dcases}
\end{equation}
where $O(1)$ represents the terms that do not scale with $D$ as $D \rightarrow \infty$. On the other hand, when the traveler is allowed to combine three means of transportation, the traveling time $T$ and resolution $E$ is given by  
\begin{align}
 T =   D/s_3 + O(1)    . 
\end{align}
It suggests that, as $D \rightarrow \infty$, the flexibility to combine walking, driving, and flying enables the traveling time to scale according to the fastest means of transportation. This phenomenon is illustrated in Fig.~S8A. We consider the setting of traveling distance $D=12$ with an accuracy requirement ranging in $W=1,2,\dots,4$. There are two means of transportation: one has speed $2.5$ with resolution $1$, the other has speed $5$ with resolution $1.5$. We consider two cases: when only one means of transportation can be used (denote as the uniform case) and when diverse means of transportation can be used (denote as the diverse case). In the diverse case, the time loss in switching between different means of transportation is $1$. In the uniform case, only $v=1,e=1$ can meet the requirement of the task with $W=1$. The error bar illustrates the upper lower bounds of the reaching time.

Compared the travel time and required accuracy for the diverse and uniform cases, Fig.~S8A shows that the diverse case performs much better than the uniform case because diverse case can exploit faster means of transportation to travel quickly while using the slower means of transportation to achieve the required accuracy, which is well in line with experimental results in Fig.~S8B.

\setcounter{figure}{0}
\makeatletter 
\renewcommand{\thefigure}{S\@arabic\c@figure}
\makeatother

\begin{figure}
 \centering
\includegraphics[width=1\textwidth]{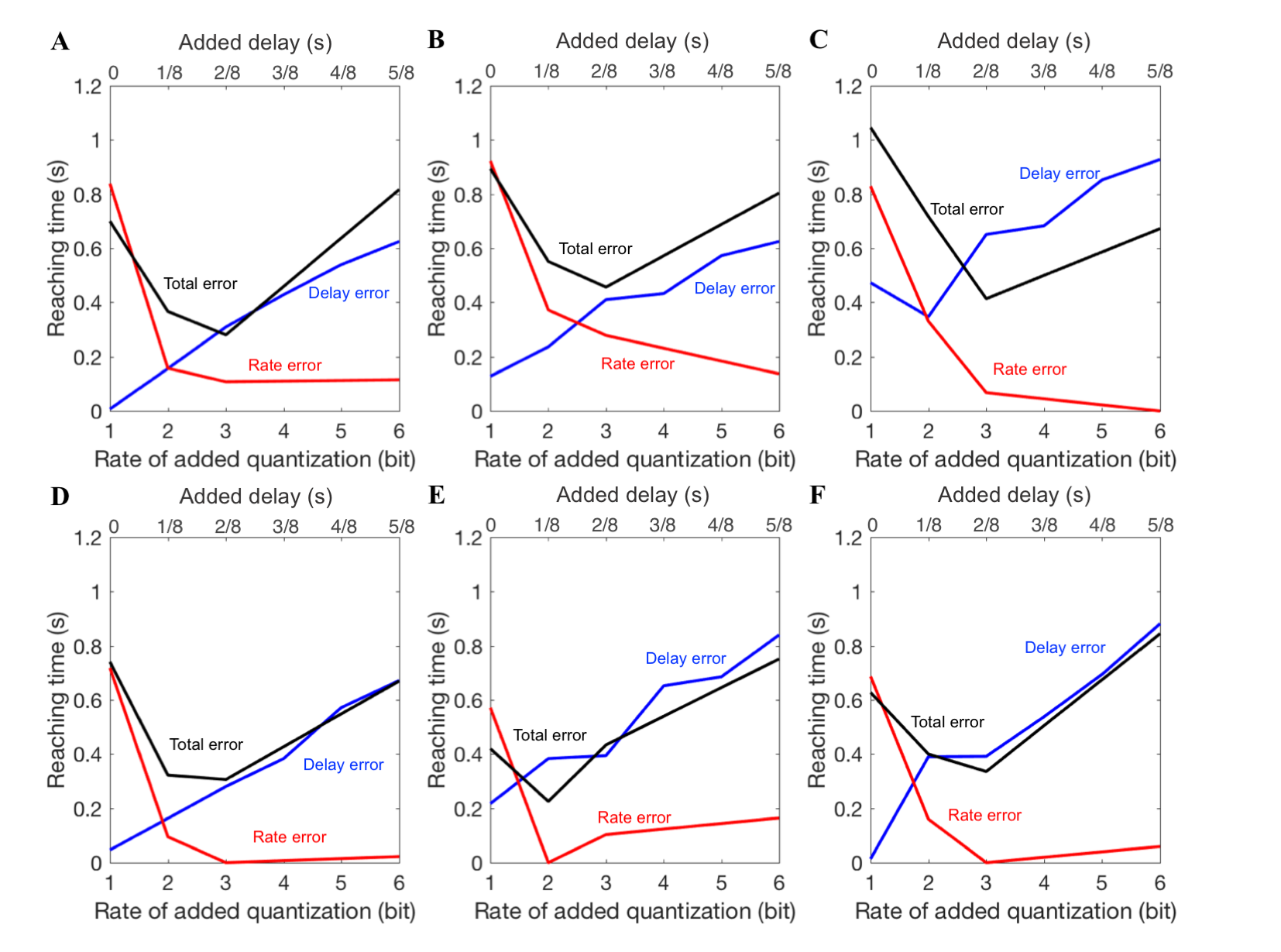}
\caption{Empirical SATs in the reaching task for each subject. The blue lines show the reaching time with added delay; the red lines show the reaching time with added quantization; and the black lines show the reaching time with both added delay and quantization, where the added delay and quantization satisfy $T  = (R-1)/8$.
}
\end{figure}

 \begin{figure}
 \centering

\begin{flushleft}
  \textbf{A}
\end{flushleft}
\center
\includegraphics[width=0.6\textwidth]{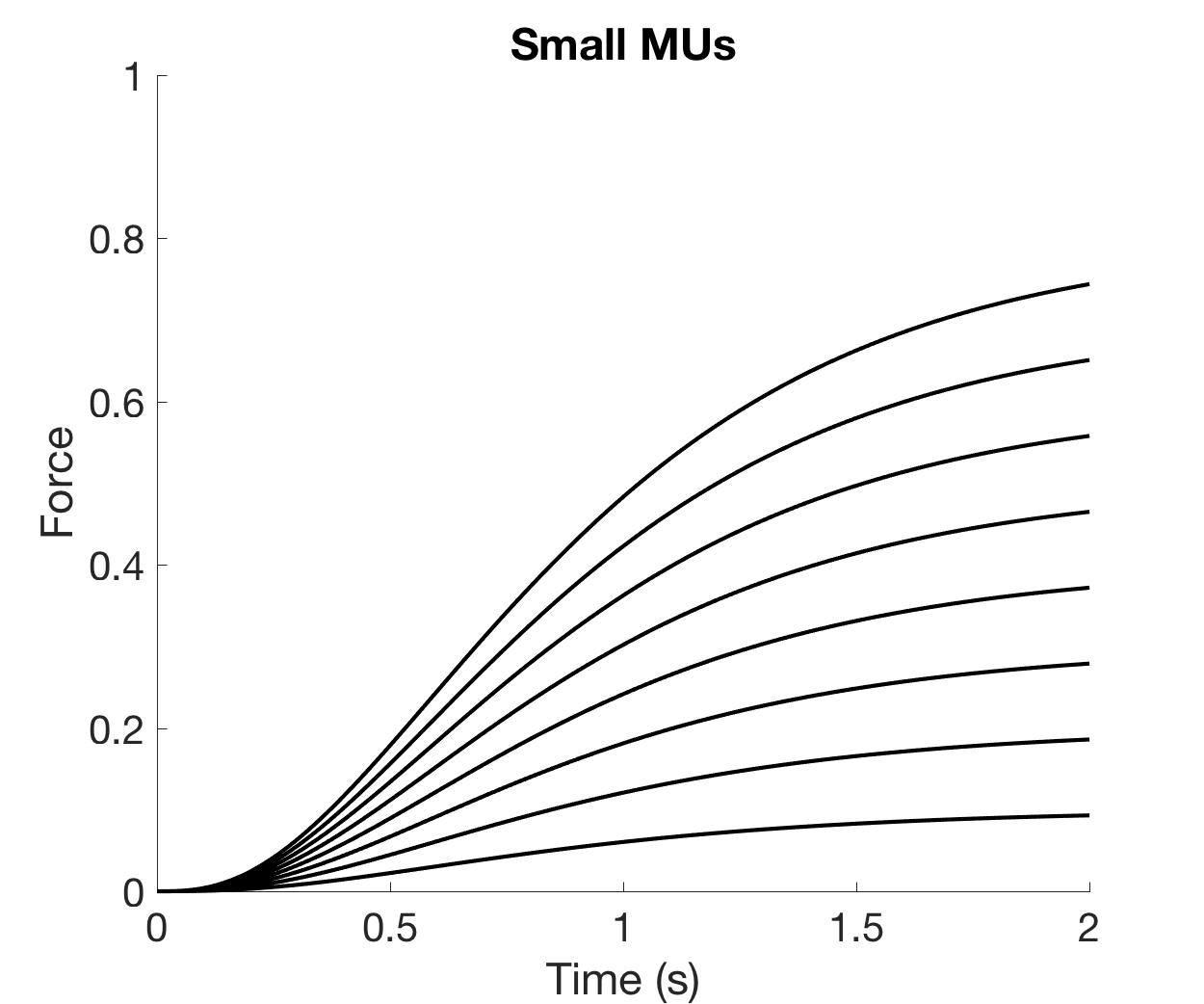}

\begin{flushleft}
  \textbf{B}
\end{flushleft}
\center
\includegraphics[width=0.6\textwidth]{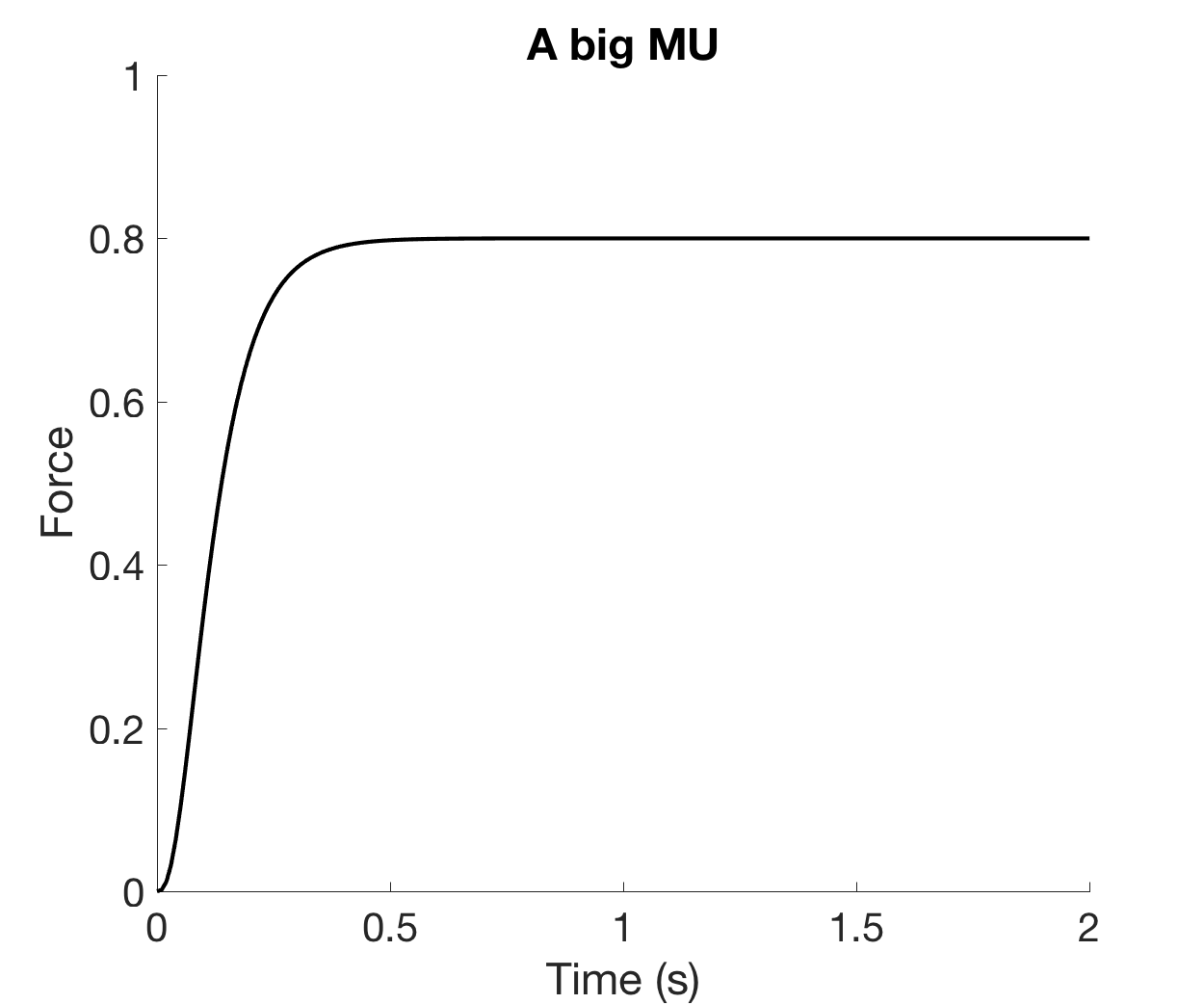}

\caption{Explanation for component-level SATs in muscles. The dashed lines show the individual and cumulative force of a group of motor units, generated from Eq. \eqref{eq:muscle-b-pde}. (A) The force from having $8$ small motor units of strength level $0.1$. (B) The response from having one large motor unit of strength level $0.8$. The total strength of all motor units for (A) and (B) are set to be same. 
}
\label{fig:muscle-SAT}
\end{figure}

\begin{figure}
 \centering

\begin{flushleft}
  \textbf{A}
\end{flushleft}
\center
\includegraphics[width=0.65\textwidth]{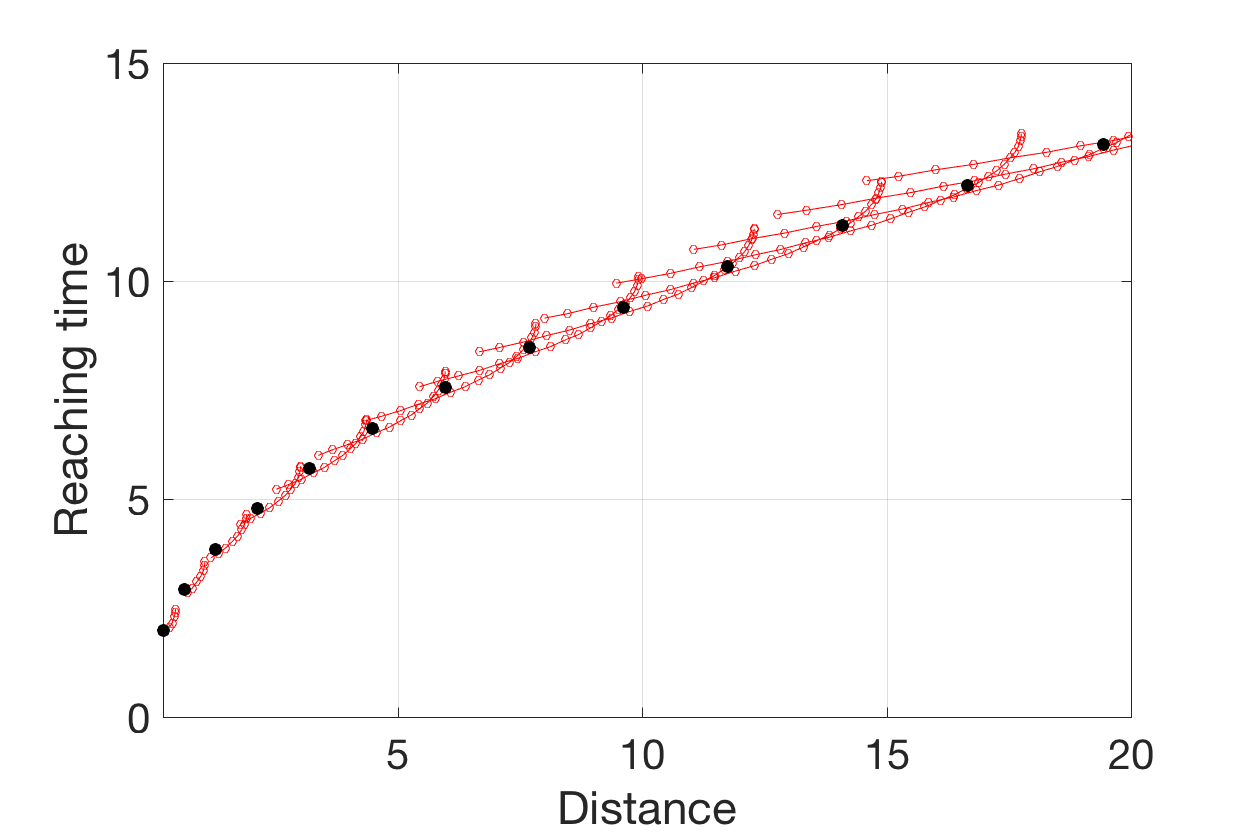}

\begin{flushleft}
  \textbf{B}
\end{flushleft}
\center
\includegraphics[width=0.6\textwidth]{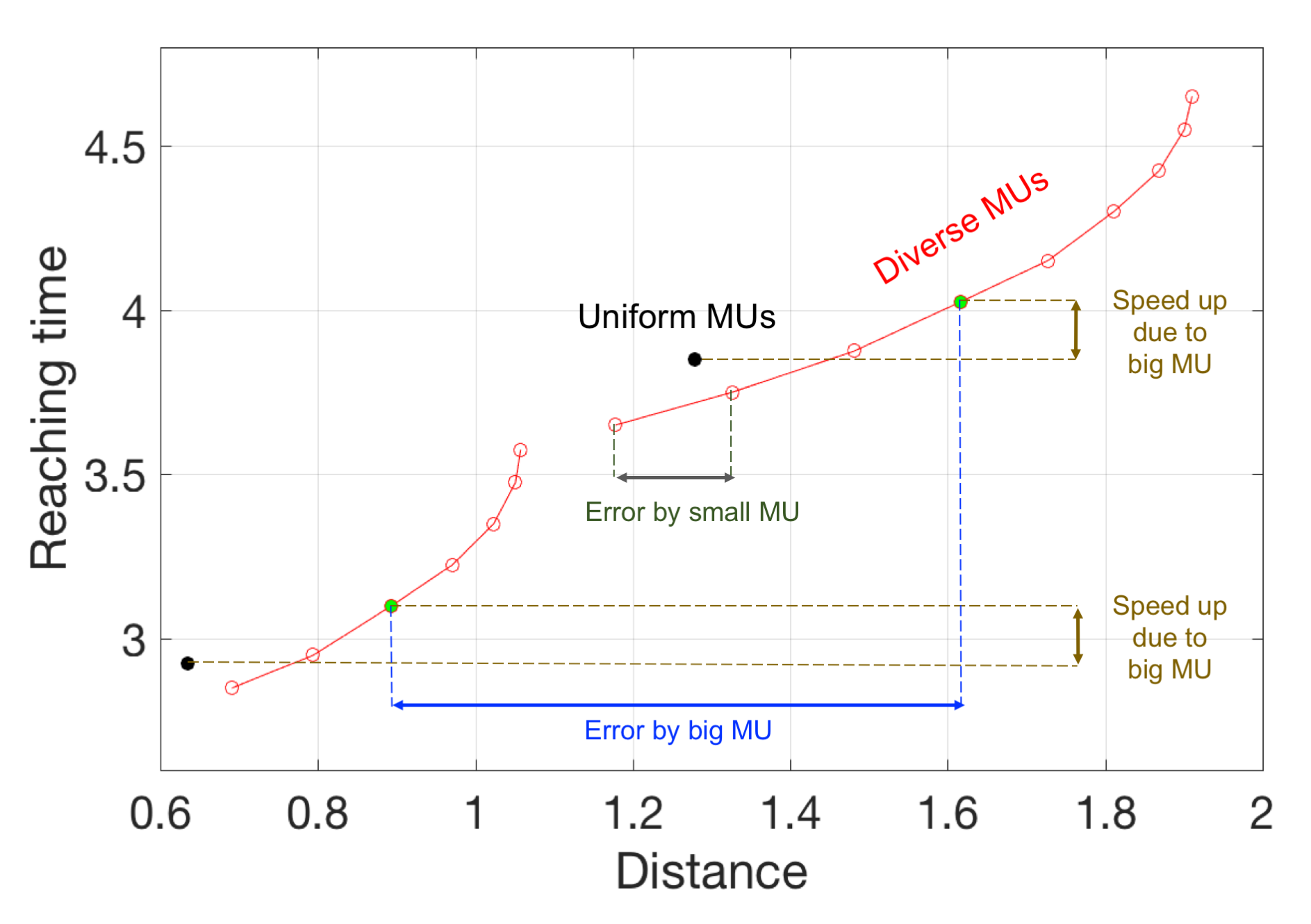}
\label{fig:muscle-system-SAT}
\caption{Achievable reaching distance and reaching time for uniform/uniform motor units in one muscle. The uniform case has two motor units with same strength ($F_1=F_2=0.5$), and the diverse case has two motor units with different strengths ($F_1=0.85, F_2=0.15$). The maximum strength that can be produced by all motor units is set to be identical in both cases. The muscle contraction duration for each motor unit ranges from $0.75$ to $14.75$ with $0.5$ increment. The friction parameters are set to be $h_s = 0.6$ and $h_k = 0.54$. Each dot represents an achievable pair between the reaching distance and time in uniform case (black) and diverse case (red). The pair with identical contraction duration in both motor units of the diverse case is colored in green. (A) Original plot. (B) Zoomed plot. 
}
\end{figure}

 \begin{figure}
 \centering

\includegraphics[width=0.8\textwidth]{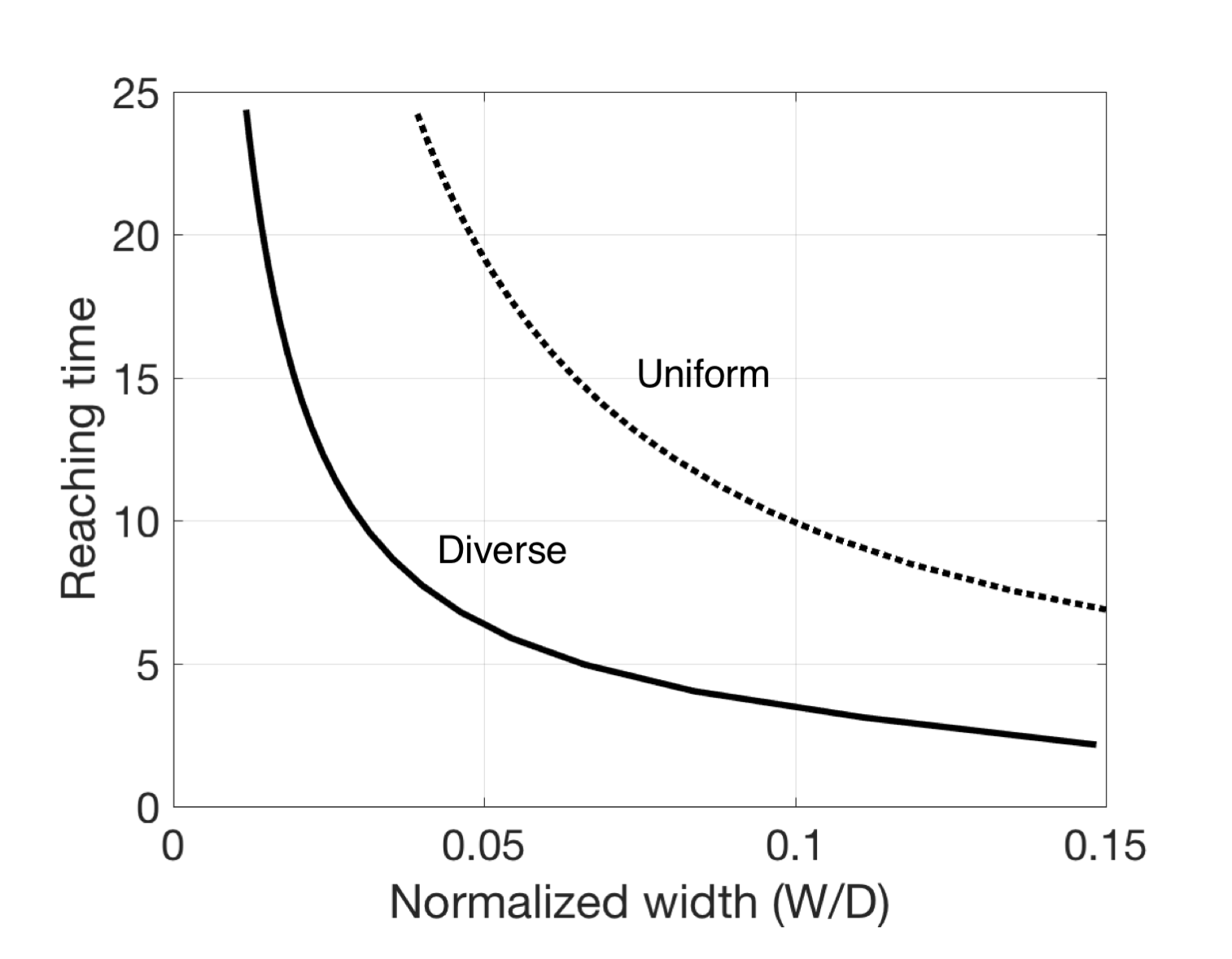}

\label{fig:muscle-system-SAT-two}
\caption{Reaching SATs when uniform or diverse muscles are used. The uniform case has two muscles with the same strength ($F_1=F_2=0.5$), and the diverse case has two muscles with different strengths ($F_1=0.85$, $F_2=0.15$). The sum of the strength of all muscles is set to be same for both cases. The friction parameters are set to be $h_s = 0.6$ and $h_k = 0.54$. 
}

\end{figure}

\begin{figure}
 \centering
\includegraphics[width=0.5\textwidth]{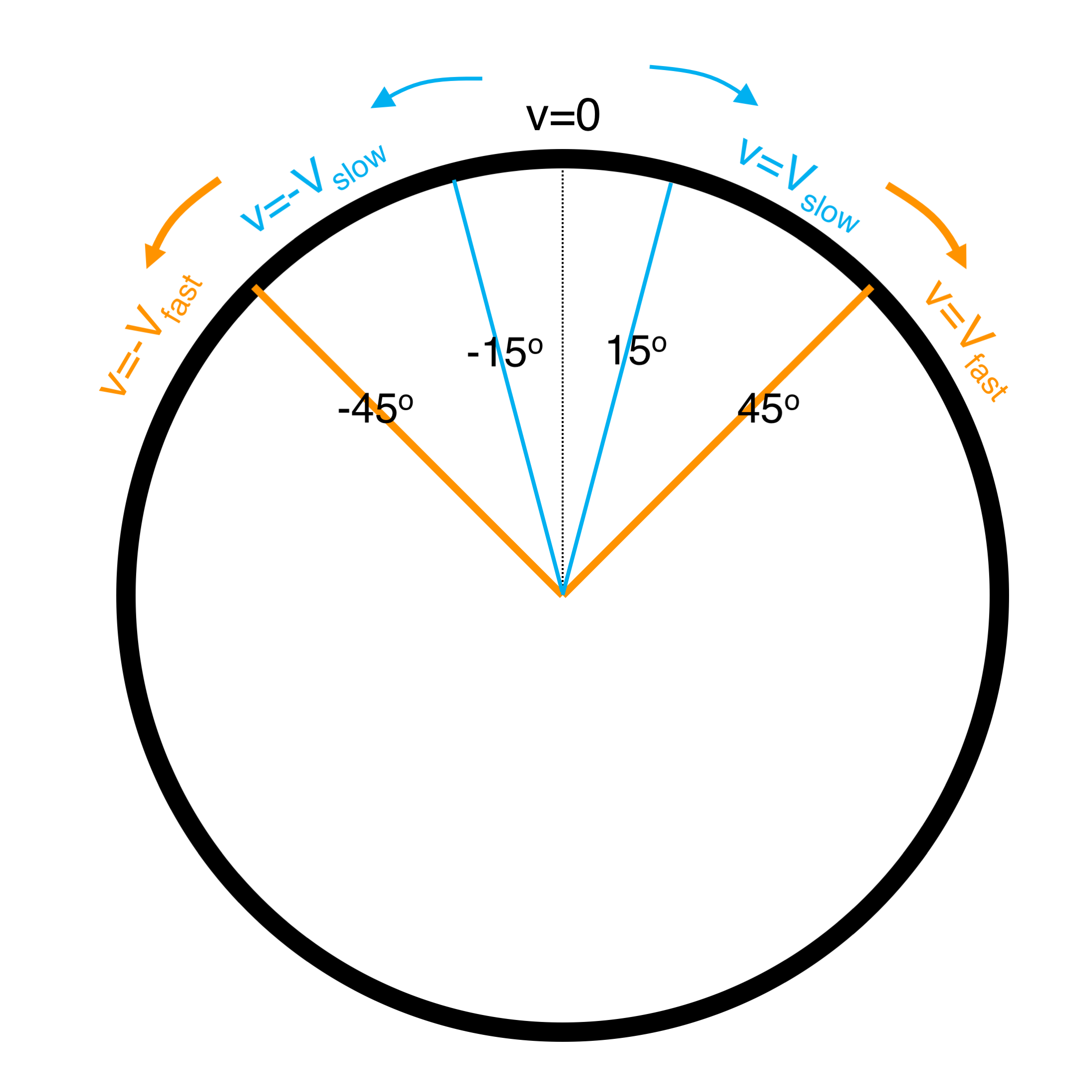}

\label{fig:muscle_2_speeds}
\caption{The angle-to-speed relation in reaching task with uniformly slow speed, uniformly fast speed, and both.}
\end{figure}

\begin{figure}
 \centering
\includegraphics[width=\textwidth]{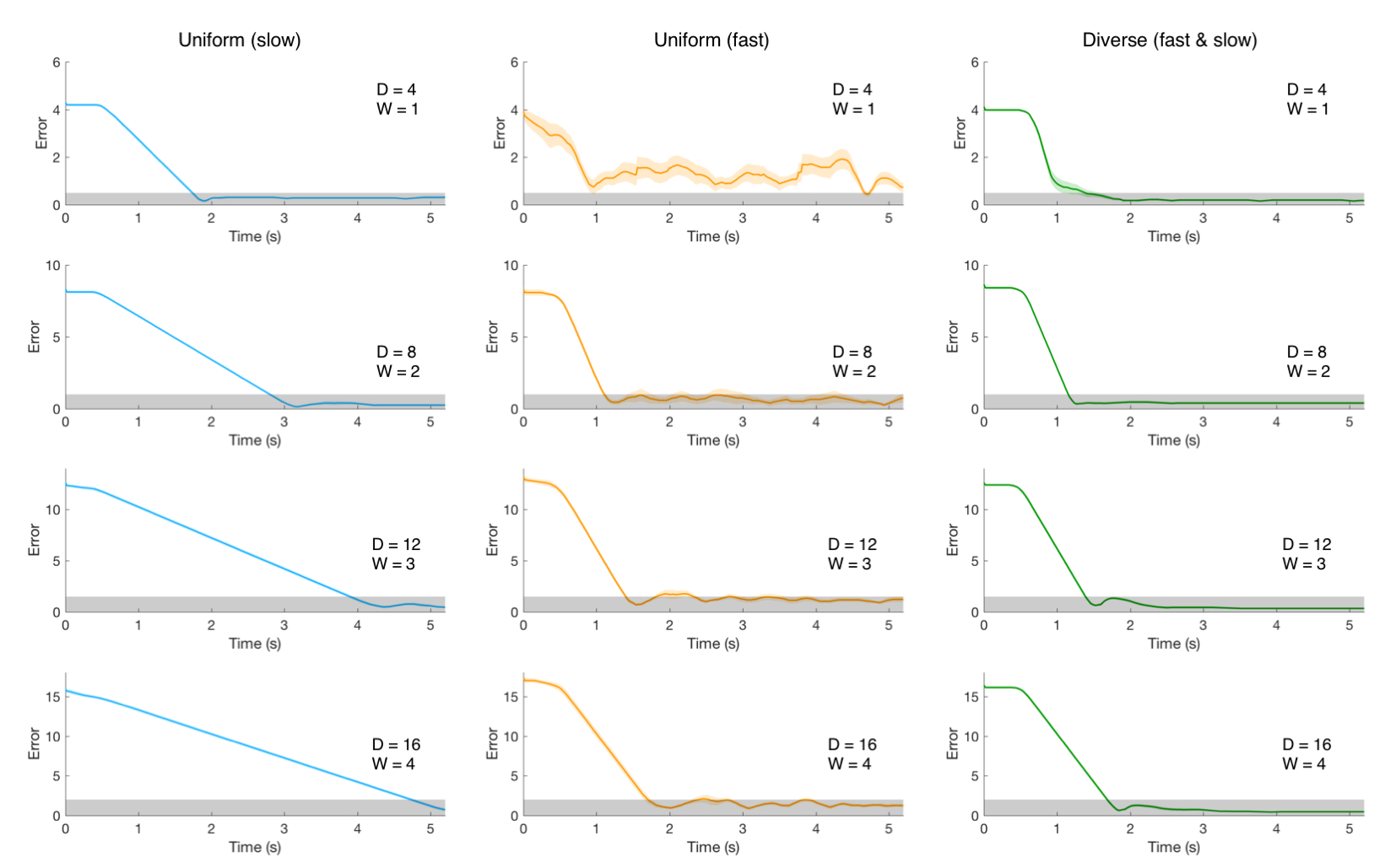}

\label{fig:muscle_2_speeds}
\caption{Error dynamics in reaching task. Blue, orange and green lines show the average error dynamics $|x(t)|$ across trials in cases with uniformly slow speed, uniformly fast speed, and both. The colored shadow shows standard error (SE). The gray zone represents the region of $|x(t)| \leq W/2$, where $W$ is the target width.}
\end{figure}

\begin{figure}
 \centering
\includegraphics[width=\textwidth]{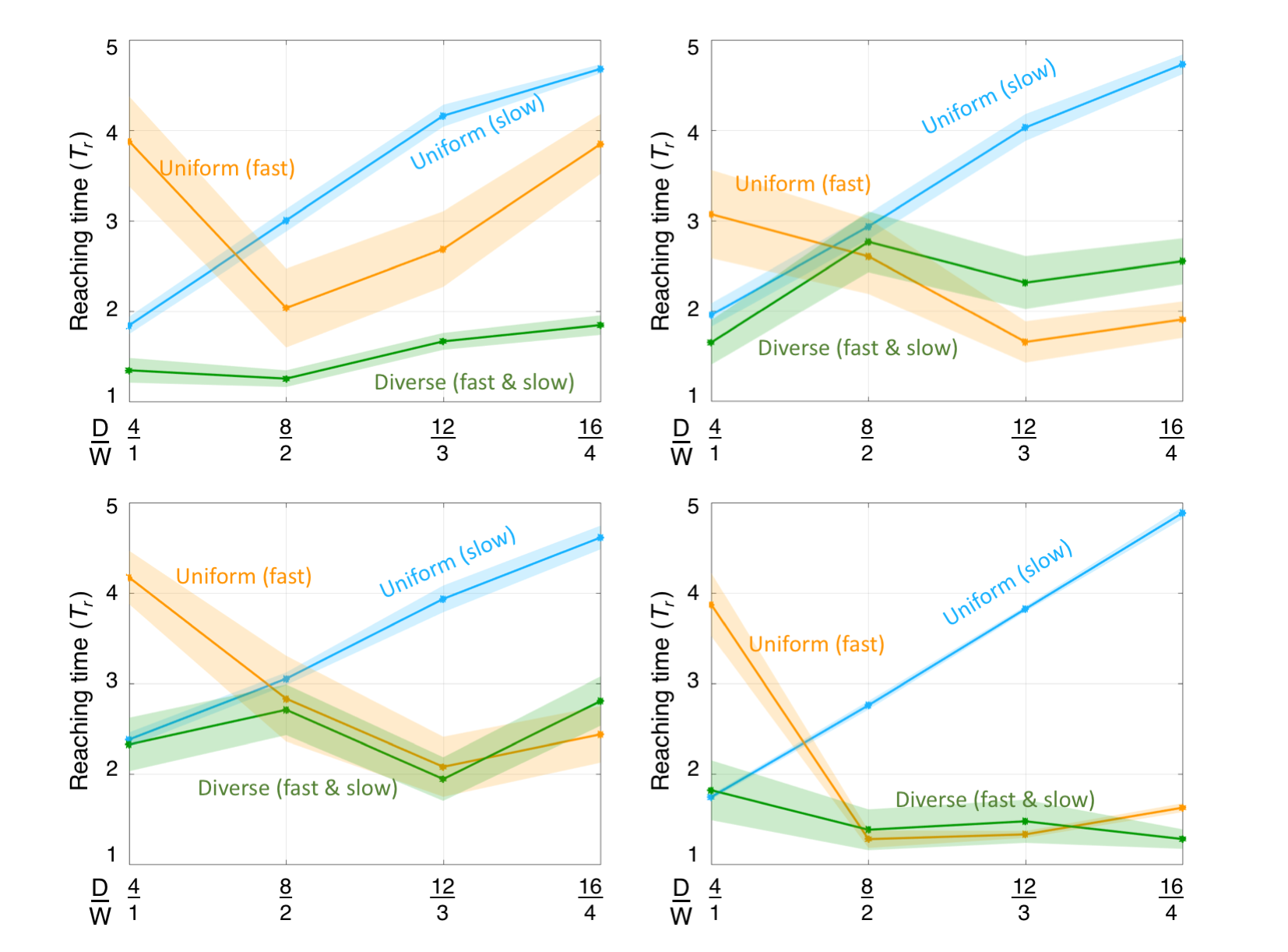}

\label{fig:muscle_exp_each_sub}
\caption{Comparisons between diverse speeds and uniform speed at single subject level. The plot shows the performance for each single subject who performed the reaching task with uniformly fast speed (orange line), uniformly slow speed (blue line), and both fast and slow speeds (green line). The standard errors across trials are shown in the shaded region around the solid lines. The results are reliable at single subject level.}
\end{figure}

\begin{figure}
\centering
\begin{flushleft}
  \textbf{A}
\end{flushleft}
\center
\includegraphics[width=0.6\textwidth]{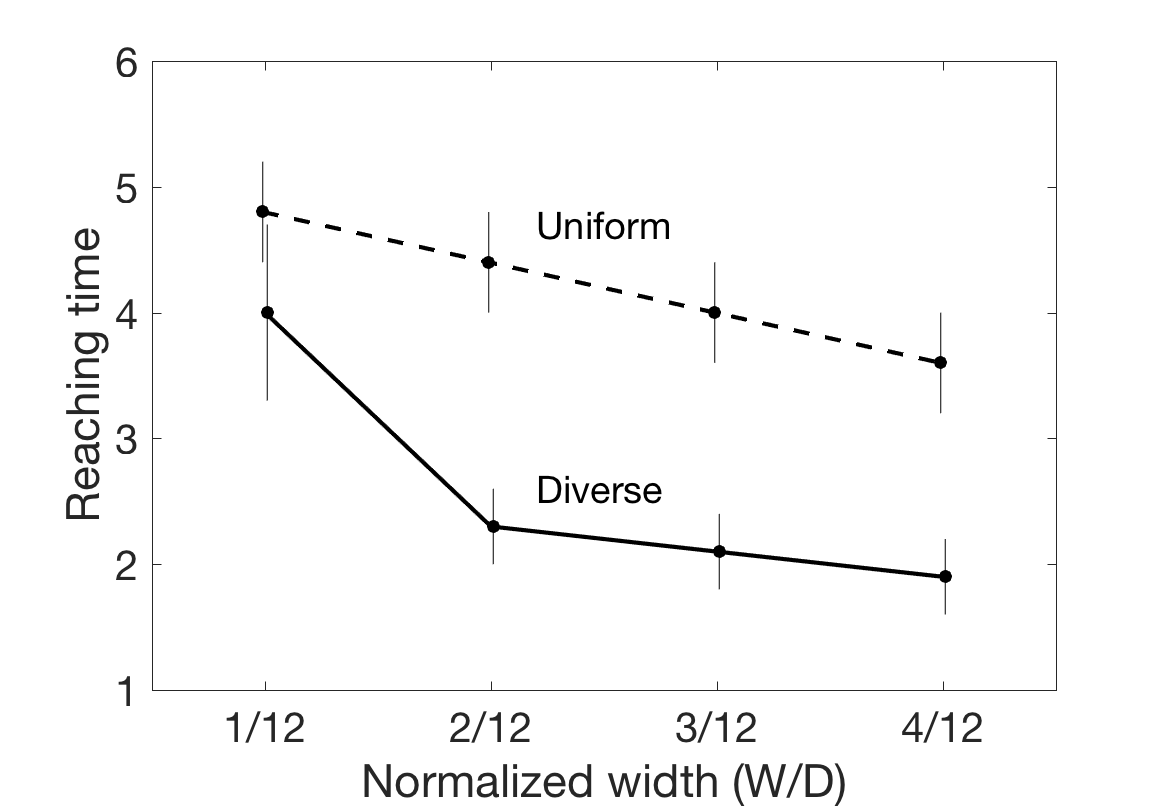}

\begin{flushleft}
  \textbf{B}
\end{flushleft}
\center
\includegraphics[width=0.6\textwidth]{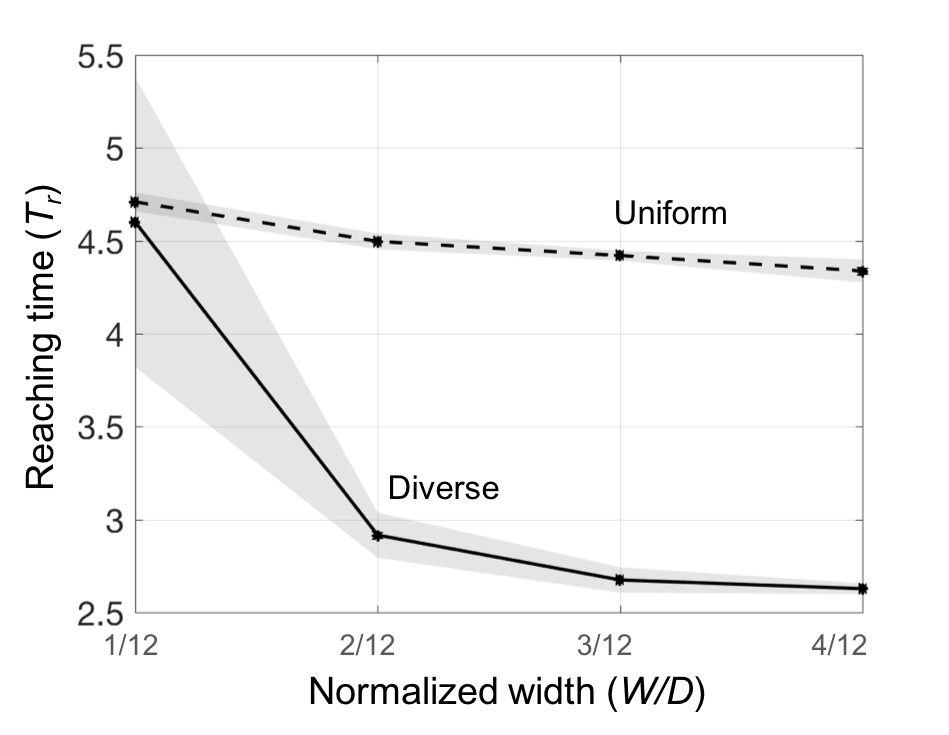}
\caption{
Reaching time versus accuracy constraints. (A) Theoretical SAT in transportation with uniform or diverse speeds.  (B) Reaching experiment result when subjects are allowed to use uniform or diverse speeds. 
}
\label{fig:transport-error}
\end{figure}

\setcounter{table}{0}
\makeatletter 
\renewcommand{\thetable}{S\@arabic\c@table}
\makeatother

\begin{table}
\begin{center}
\begin{tabular}{@{\extracolsep{5pt}}ll} 
\\[-1.8ex]\hline 
\hline \\[-1.8ex] 
\textbf{Parameter} & \textbf{Description} \\
\hline 
$x(t)$ & Error at time step $t$  \\
\hline
$\mathcal K$ & Controller  \\
\hline
$T_s \geq 0$ & Signaling delay  \\
\hline
$T_i \geq 0$ & Internal delay \\
\hline
$T = T_s + T_i$ & Total delay 
\\
\hline
$T_t$ & Time to reach target \\
\hline
$R$ & Information rate (bits per unit time) 
\\
\hline
$\lambda$ & Cost associated with the resource use
\\
\hline
\end{tabular}
\caption{Parameters in the basic model.
}
\label{tab:notations}
\end{center}
\end{table}


\bibliography{scibib}

\begin{thebibliography}{10}
\bibitem{big_animal}
Bigger animals have more inertia and can tolerate longer delays. they also use
  more time to compute in part because large animals (elephants, etc.) are more
  prone to falling.
\bibitem{note1f}
In other words, the time required for a muscle to reach to $c_i(t) = f_i$ from
  $c_i(0) = 0$ is decreasing in $f_i$.
\bibitem{note4f}
The Kronecker delta function is defined as follows: $\delta (t) = 1$ if $t =
  0$, and $\delta (t) = 0$ otherwise.
\end{thebibliography}
\begin{thebibliography}{10}
\bibitem{log-laws}
These laws include the weber-fechner law, a log relation between the physical
  change in a stimulus and the perceived change in human perception; the ricco
  law for visual target detection for unresolved targets; the accot-zhai law
  for steering, a generalization of fitts' law for 2d environments; the spacing
  effect of ebbinghaus for long-term recall from memory; and the hick-hyman law
  for the logarithmic increase in the time it takes to make a decision as the
  number of choices increases.
\end{thebibliography}
\bibliographystyle{Science}

\end{document}